\documentstyle[epsfig]{mn}

\newif\ifAMStwofonts

\ifoldfss
  \ifCUPmtlplainloaded \else
    \NewTextAlphabet{textbfit} {cmbxti10} {}
    \NewTextAlphabet{textbfss} {cmssbx10} {}
    \NewMathAlphabet{mathbfit} {cmbxti10} {} 
    \NewMathAlphabet{mathbfss} {cmssbx10} {} 
  \fi
  \ifAMStwofonts
    \ifCUPmtlplainloaded \else
      \NewSymbolFont{upmath} {eurm10}
      \NewSymbolFont{AMSa} {msam10}
      \NewMathSymbol{\upi}     {0}{upmath}{19}
      \NewMathSymbol{\umu}     {0}{upmath}{16}
      \NewMathSymbol{\upartial}{0}{upmath}{40}
      \NewMathSymbol{\leqslant}{3}{AMSa}{36}
      \NewMathSymbol{\geqslant}{3}{AMSa}{3E}

      \let\leq=\leqslant 
      \let\geq=\geqslant 
    \fi
  \fi
\fi 

\ifnfssone
  \newmathalphabet{\mathit}
  \addtoversion{normal}{\mathit}{cmr}{m}{it}
  \addtoversion{bold}{\mathit}{cmr}{bx}{it}
  \newmathalphabet{\mathbfit} 
  \addtoversion{normal}{\mathbfit}{cmr}{bx}{it}
  \addtoversion{bold}{\mathbfit}{cmr}{bx}{it}
  \newmathalphabet{\mathbfss} 
  \addtoversion{normal}{\mathbfss}{cmss}{bx}{n}
  \addtoversion{bold}{\mathbfss}{cmss}{bx}{n}
  \ifAMStwofonts
    \ifCUPmtlplainloaded \else
      \UseAMStwoboldmath
      \makeatletter
      \new@mathgroup\upmath@group
      \define@mathgroup\mv@normal\upmath@group{eur}{m}{n}
      \define@mathgroup\mv@bold\upmath@group{eur}{b}{n}
      \edef\UPM{\hexnumber\upmath@group}
      \new@mathgroup\amsa@group
      \define@mathgroup\mv@normal\amsa@group{msa}{m}{n}
      \define@mathgroup\mv@bold\amsa@group{msa}{m}{n}
      \edef\AMSa{\hexnumber\amsa@group}
      \makeatother
      \mathchardef\upi="0\UPM19
      \mathchardef\umu="0\UPM16
      \mathchardef\upartial="0\UPM40
      \mathchardef\leqslant="3\AMSa36
      \mathchardef\geqslant="3\AMSa3E

      \let\leq=\leqslant 
      \let\geq=\geqslant 
    \fi
  \fi
\fi 

\ifnfsstwo
  \DeclareMathAlphabet{\mathbfit}{OT1}{cmr}{bx}{it}
  \SetMathAlphabet\mathbfit{bold}{OT1}{cmr}{bx}{it}
  \DeclareMathAlphabet{\mathbfss}{OT1}{cmss}{bx}{n}
  \SetMathAlphabet\mathbfss{bold}{OT1}{cmss}{bx}{n}
  \ifAMStwofonts
    \ifCUPmtlplainloaded \else
      \DeclareSymbolFont{UPM}{U}{eur}{m}{n}
      \SetSymbolFont{UPM}{bold}{U}{eur}{b}{n}
      \DeclareSymbolFont{AMSa}{U}{msa}{m}{n}
      \DeclareMathSymbol{\upi}{0}{UPM}{"19}
      \DeclareMathSymbol{\umu}{0}{UPM}{"16}
      \DeclareMathSymbol{\upartial}{0}{UPM}{"40}
      \DeclareMathSymbol{\leqslant}{3}{AMSa}{"36}
      \DeclareMathSymbol{\geqslant}{3}{AMSa}{"3E}

      \let\leq=\leqslant 
      \let\geq=\geqslant 
    \fi
  \fi
\fi 

\ifCUPmtlplainloaded \else
  \ifAMStwofonts \else 
    \def\upi{\pi}
    \def\umu{\mu}
    \def\upartial{\partial}
  \fi
\fi

\title[HST imaging of hyperluminous infrared galaxies]
  {HST/WFPC2 observations of hyperluminous infrared galaxies}
\author[D. Farrah et al.]
  {D.~Farrah$^1$, A.~Verma$^2$, S.~Oliver$^3$, M.~Rowan-Robinson$^1$, R.~McMahon$^4$\\
 $^1$Blackett Laboratory, Imperial College, Prince Consort Road, London SW7 2BW, UK\\
 $^2$Max-Planck-Institut fur Extraterrestrische Physik, Postfach 1312, 85741 Garching, Germany\\
 $^3$Astronomy Centre, University of Sussex, Falmer, Brighton BN1 9QJ, UK\\
 $^4$Institute of Astronomy, University of Cambridge, Madingley Road, Cambridge CB3 0HA, UK\\
}
\date{Received 2001 April 30}
\pagerange{\pageref{01}--\pageref{11}}
\pubyear{2001}

\def\LaTeX{L\kern-.36em\raise.3ex\hbox{a}\kern-.15em
    T\kern-.1667em\lower.7ex\hbox{E}\kern-.125emX}

\begin{document}

\label{firstpage}

\maketitle

\begin{abstract}
We present {\em HST} WFPC2 {\em I} band imaging for a sample of 9 Hyperluminous 
Infrared Galaxies spanning a redshift 
range $0.45 < z < 1.34$. Three of the sample have morphologies showing evidence for 
interactions, six are QSOs. Host galaxies in the QSOs are reliably detected out to 
$z \sim 0.8$. The detected QSO host galaxies have an elliptical morphology with 
scalelengths spanning $6.5 < r_{e}(Kpc) < 88$ and absolute $k$ corrected magnitudes 
spanning $-24.5 < M_{I} < -25.2$. There is no clear correlation between the IR power source 
and the optical morphology. None of the sources in the sample, including F15307+3252, show 
any evidence for gravitational lensing. We infer that the IR luminosities are thus real.
Based on these results, and previous studies of HLIRGs, we conclude that this class of object 
is broadly consistent with being a simple extrapolation of the ULIRG population to higher 
luminosities; ULIRGs being mainly violently interacting systems powered by starbursts and/or 
AGN. Only a small number of sources whose infrared luminosities exceed 10$^{13}L_{\sun}$ are intrinsically 
less luminous objects which have been boosted by gravitational lensing.\end{abstract}


\section{Introduction}
One of the most important results from the Infrared Astronomical Satellite ({\em IRAS}) all 
sky surveys was the detection of a new class of galaxy where the bulk of the bolometric emission 
lies in the infrared range \cite{so1,sa1}. This population becomes the dominant extragalactic 
population at luminosities above $L_{IR}>10^{11}L_{\sun}$. At the extreme upper end of the 
{\em IRAS} galaxy population lie the Hyperluminous Infrared Galaxies (HLIRGs), those with 
$L_{IR} > 10^{13.0} h_{65}^{-2} L_{\sun}$ \cite{rr2}. The first HLIRG to be found, P09104+4109, was 
identified by Kleinmann et al \shortcite{kl}, with a far infrared luminosity of 
$1.5\times10^{13}h_{50}^{-2}L_{\sun}$. In 1991, Rowan-Robinson et al identified F10214+4724 with 
$z=2.286$ and a far infrared luminosity of $3\times10^{14}h^{-2}_{50}L_{\sun}$. Later observations 
of this object revealed a huge mass of molecular gas ($10^{11}h_{50}^{-2}M_{\sun}$ 
\cite{br,so3}), a Seyfert emission spectrum \cite{el}, very high optical polarisation \cite{la}, and 
strong evidence for lensing with a magnification of about 10 at infrared wavelengths \cite{gra,bro,ser,ei,gr2}. 
These objects appeared to presage an entirely new class of infrared galaxy.

There are currently about 50 sources positively identified as HLIRGs, of which 13 were identified in 
spectroscopic follow-up observations of sources discovered in far infrared or sub-mm surveys \cite{rr2}. Two of those 13 IR-selected 
HLIRGs have been found to be lensed. By projecting population densities from the {\em IRAS} Faint Source Survey 
Rowan-Robinson \shortcite{rr2} estimates that there are between 100 and 200 HLIRGs with 
60$\mu$m flux  $>$ 200 mJy over the whole sky.

\begin{table*}
\begin{minipage}{175mm}
\caption{Hyperluminous Galaxies Observed By {\em HST} \label{hstobs}}
\begin{tabular}{@{}ccccccccccc}
Name         & RA (2000)  & Dec        & $z$  &$f_{814}^1$& $m_{I}$ &$M_{I}^2$& $L_{IR}^3$& IR SED$^7$            & Opt Spec.$^8$& Morphology  \\
F00235+1024  & 00 26 06.7 & 10 41 27.6 & 0.58 & 0.08      & 18.73   &-23.6    & 13.1$^4$  & 63\% Sb, 37\% AGN$^4$ & NL           & Interacting \\   
F10026+4949  & 10 05 52.5 & 49 34 47.8 & 1.12 & 0.04      & 19.61   &-24.3    & 13.8$^5$  & AGN dominated$^5$     & Sy1          & Interacting \\   
PG1148+549   & 11 51 20.4 & 54 37 32.8 & 0.97 & 2.07      & 15.21   &-28.3    & 13.7      & 30\% Sb, 70\% AGN     & QSO          & Undisturbed \\
LBQS1220+0939& 12 23 17.9 & 09 23 07.3 & 0.68 & 0.43      & 16.91   &-25.8    & 13.5$^5$  & -                     & QSO          & Undisturbed \\
F12358+1807  & 12 38 20.2 & 17 50 38.8 & 0.45 & 1.09      & 15.90   &-25.9    & 12.7      & -                     & BAL QSO      & Undisturbed \\
F12509+3122  & 12 53 17.6 & 31 05 50.5 & 0.78 & 0.91      & 16.10   &-26.9    & 13.5$^5$  & AGN dominated$^5$     & QSO          & Undisturbed \\
F14218+3845  & 14 23 55.5 & 38 31 51.3 & 1.21 & 0.07      & 18.96   &-25.1    & 13.1$^4$  & 74\% Sb, 26\% AGN$^4$ & QSO          & Undisturbed \\
F15307+3252  & 15 32 44.1 & 32 42 46.2 & 0.93 & 0.14      & 18.20   &-25.2    & 13.5$^4$  & 32\% Sb, 68\% AGN$^4$ & Sy2          & Interacting \\   
PG1634+706   & 16 34 28.8 & 70 31 32.8 & 1.33 & 6.30      & 14.00   &-30.4    & 14.0      & 17\% Sb, 83\% AGN     & QSO          & Undisturbed \\
\end{tabular}

\medskip
Coordinates and absolute {\em F814W} band magnitudes/fluxes are taken from the {\em HST} images and are in the 
V{\sevensize EGAMAG} {\em HST} flight filter system.
$^1${\em I} band flux in units of mJy. Subscript denotes wavelength in microns.
$^2$No applied {\em k} correction. 
$^3$Logarithm of the $1 - 1000\mu$m IR luminosity, where necessary recomputed for $H_{0}=65$ km s$^{-1}$ Mpc$^{-1}$ and 
$\Omega_{0}=1.0$, and in units of bolometric Solar luminosities.
$^4$Verma et al 2001. 
$^5$Rowan-Robinson 2000. 
$^6$Haas et al 1998. 
$^7$IR Spectral Energy Distribution, computed between $1-1000\mu$m. 
$^8$Optical spectral classification, taken from Rowan-Robinson 2000.
\end{minipage}
\end{table*}

The source and trigger of the IR emission in HLIRGs is currently the subject of considerable 
debate. ULIRGs appear to be powered by starburst and/or quasar activity triggered by interactions \cite{le,cl}, 
the HLIRGs may simply be the high luminosity tail of the ULIRG population. Rowan-Robinson 
\shortcite{rr2} argues that the majority of the emission at rest-wavelengths $>$50$\mu$m in 
HLIRGs is due to starburst activity, implying star formation rates $>1000M_{\sun}yr^{-1}$. If the 
rest-frame far infrared and sub-mm emission from HLIRGs is due to star formation, then the implied 
star formation rates would be the highest for any objects in the Universe. This would strongly 
suggest these galaxies are going through their maximal star formation periods, implying that they 
are very young galaxies. A third possibility is that if the IR emission arises via some other 
mechanism (e.g. a transient IR luminous phase in QSO evolution) then these galaxies may represent 
an entirely different class of object.

In this paper we examine the morphologies of HLIRGs, their host galaxies and immediate environments 
in an attempt to discern the nature and origins of their IR emission. We present a sample of 
Hyperluminous Infrared Galaxies imaged with the Wide Field Planetary Camera 2 on board the Hubble 
Space Telescope. Sample selection and observations, and data analysis are described in \S\S 2 and 
3. Results are presented in \S 4, including sample morphology, 
lensing properties and descriptions of individual sources. Discussion of these results can be 
found in \S 5 and conclusions can be found in \S 6.   

Unless otherwise stated we have taken $H_{0}=65$ km s$^{-1}$ Mpc$^{-1}$ and $\Omega_{0}=1.0$.

\section{Observations}
\subsection{Sample}
The 9 objects in this sample all have a far infrared luminosity close to or exceeding $10^{13} L_{\sun}$ that cannot be 
explained by non-thermal emission. The redshift range spans $z=0.44$ to $z=1.34$, with a mean redshift of $z=0.9$. Of the 
objects in this sample, 7 were identified purely by IR selection techniques.

Three of the sources, F10026+4949, F12509+3122 and F14218+3845 were discovered as part of a redshift survey of 3703 
{\em IRAS} FSS objects to a flux limit of 200mJy \cite{ol}. PG1148+549 and PG1634+706 are {\em IRAS} detected PG QSOs with SEDs 
indicating an IR 
luminosity of $L_{IR}>10^{13}L_{\sun}$. F00235+1024, LBQS1220+0939 and F12358+1807 were discovered as the result of a 
systematic program of statistical optical identification being carried out with the APM machine (McMahon et al 2001, in 
preparation). This project uses robust statistical estimators to identify {\em IRAS} sources taking into account the 
{\em IRAS} error ellipse and optical magnitudes of all potential optical counterparts. LBQS1220+0939 and F12358+1807 are 
associated with known quasars with $m_{B}<18.5$, i.e. 10 times less bright in the rest frame UV than the PG quasars. 
F12358+1807 lies slightly below the $10^{13}L_{\sun}$ limit but is a member of the Broad Absorption Line quasar class. 
BAL QSOs are believed to exhibit excess FIR emission and F12358+1807 is the most luminous object in its class. F15307+3252 
is an {\em IRAS} source \cite{cu} that has an IR luminosity in excess of $10^{13} L_{\sun}$.

The {\em HST} data were taken in cycle 6 between March and July 1997 using the Wide Field Planetary Camera 2. The coordinates 
of each source were centred on the Planetary Camera CCD, selected for its superior pixel scale and full sampling of 
the PSF over the Wide Field Camera CCDs (0.046\arcsec\ pixel$^{-1}$ as opposed to $\sim$ 0.1\arcsec\ pixel$^{-1}$). Exposures 
were taken using the F814W filter, which corresponds closely to the {\em I} band filter in the Cousins system. Each observation 
consisted of one short exposure (100 - 600s) to image the brightest parts of each object with minimal saturation of the 
detector, and two or three longer exposures (700 - 1600s), taken both to image the fainter regions of each source and to 
facilitate the subtraction of cosmic rays.

\section{Data Analysis}

\subsection{Data Reduction}
All datasets were calibrated on arrival with the best available reference files, using the 
{\em IRAF} task C{\sevensize ALWP2}, and combined into a single image using the {\em IRAF} 
task G{\sevensize COMBINE}. Statistical weights were assigned to different exposures based 
on their exposure times. The  G{\sevensize COMBINE} rejection algorithm C{\sevensize CDCRREJ}, 
which incorporates the Planetary Camera CCD characteristics, was used to remove cosmic rays by 
comparison between different exposures. Pixels flagged as saturated in the deeper exposures 
were replaced by their scaled unsaturated counterparts from one of the shallower exposures, with suitably 
modified statistical weights. Following these steps the sky background was subtracted and warm 
pixels were removed by linear interpolation from their immediate neighbours.

Photometric solutions were calculated using the {\em Synphot} package from the STScI, which is consistent 
with the solutions given by Holtzman et al \shortcite{hol}, and are given in the V{\sevensize EGAMAG} 
system. For the QSOs in the sample that were saturated in the central regions PSF fitting photometry was performed to accurately 
measure the true source magnitudes. Corrections were made for detector gain (for these observations the gain 
was $7e^{-}/DN$) and aperture size. Absolute magnitudes (without a {\em k} correction term) were derived 
using the expression:

\begin{equation}
M = m-25-5log[(2c/H_{0})(1+z)(1-(1+z)^{-1/2})]
\end{equation}

\noindent We estimate that our derived relative magnitudes have an associated photometric error of approximately 5\%. 
Prescriptions for converting magnitudes in the {\em HST} filter system (the `flight' system) into more conventional filter 
systems depend in some way on the object colours. The F814W filter is a good match to the Cousins {\em I} band filter, the 
conversion factor is very small ($<0.05$ magnitudes) and depends only slightly on the object colour. As 
the colours of our sample are not known we have not applied a conversion factor and our magnitudes, although referred to as 
{\em I} band magnitudes, remain in the WFPC2 flight filter system. The difference between these magnitudes and the equivalent 
Cousins {\em I} band magnitudes is small ($<0.1$ magnitudes) and thus comparisons can readily be made between the two systems. 
Fluxes in Jy were obtained by multiplying the flux in ergs cm$^{-2}$ s$^{-1}$ \AA$^{-1}$ by the central 
wavelength of the F814W filter in angstroms, and then dividing by the central frequency in Hz. As such they do not include 
any bandpass colour correction.  

{\em k} corrections for the F814W filter were computed using the spectral synthesis package P{\sevensize EGASE} and the 
associated galaxy templates \cite{fi}. Previous studies \cite{su} have shown that the optical emission from very 
luminous {\em IRAS} galaxies is dominated by emission from old stellar populations, rather than an unobscured starburst. 
We therefore used models for evolved stellar populations to calculate {\em k} corrections. The galaxy template for a stellar 
population of age $10^{9}$ years was generated using a Rana \& Basu (1992) IMF and a star formation law $\tau(t) = \nu g(t)$ 
where $g(t)$ is the gas fraction. Published values of {\em k} corrections can vary markedly, especially for bluer filters 
(e.g. Poggianti 1997). We estimate that the error in our computed values due to uncertainties in Initial Mass Functions, 
star formation laws, the age of the system and the assumed source spectrum is approximately 0.2 magnitudes at $z = 0.45$, 
rising to 0.45 magnitudes at $z = 1.33$. No corrections have been made for reddening due to dust extinction.

\begin{figure*}
\begin{minipage}{170mm}
\epsfig{figure=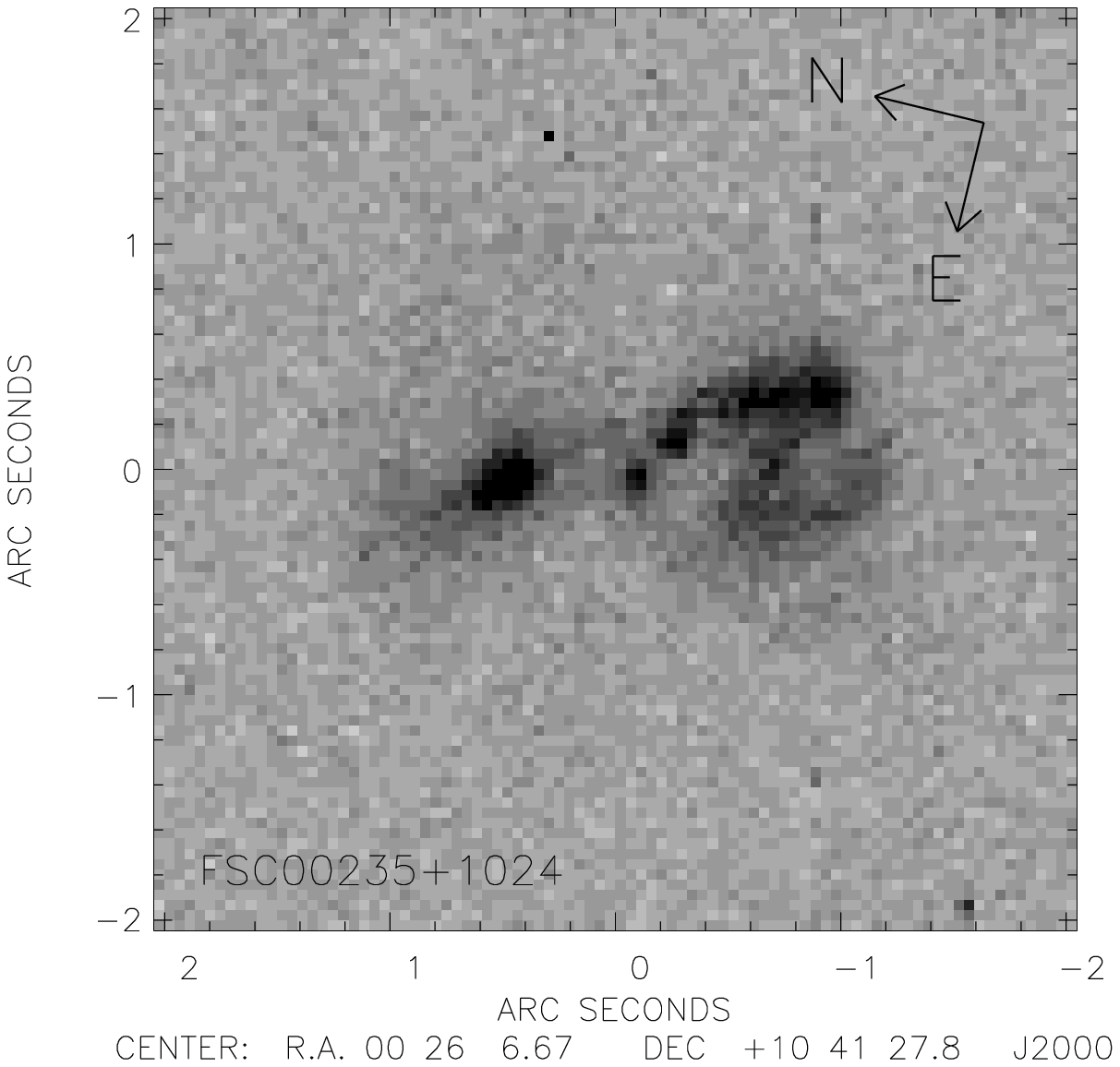,width=55mm}
\epsfig{figure=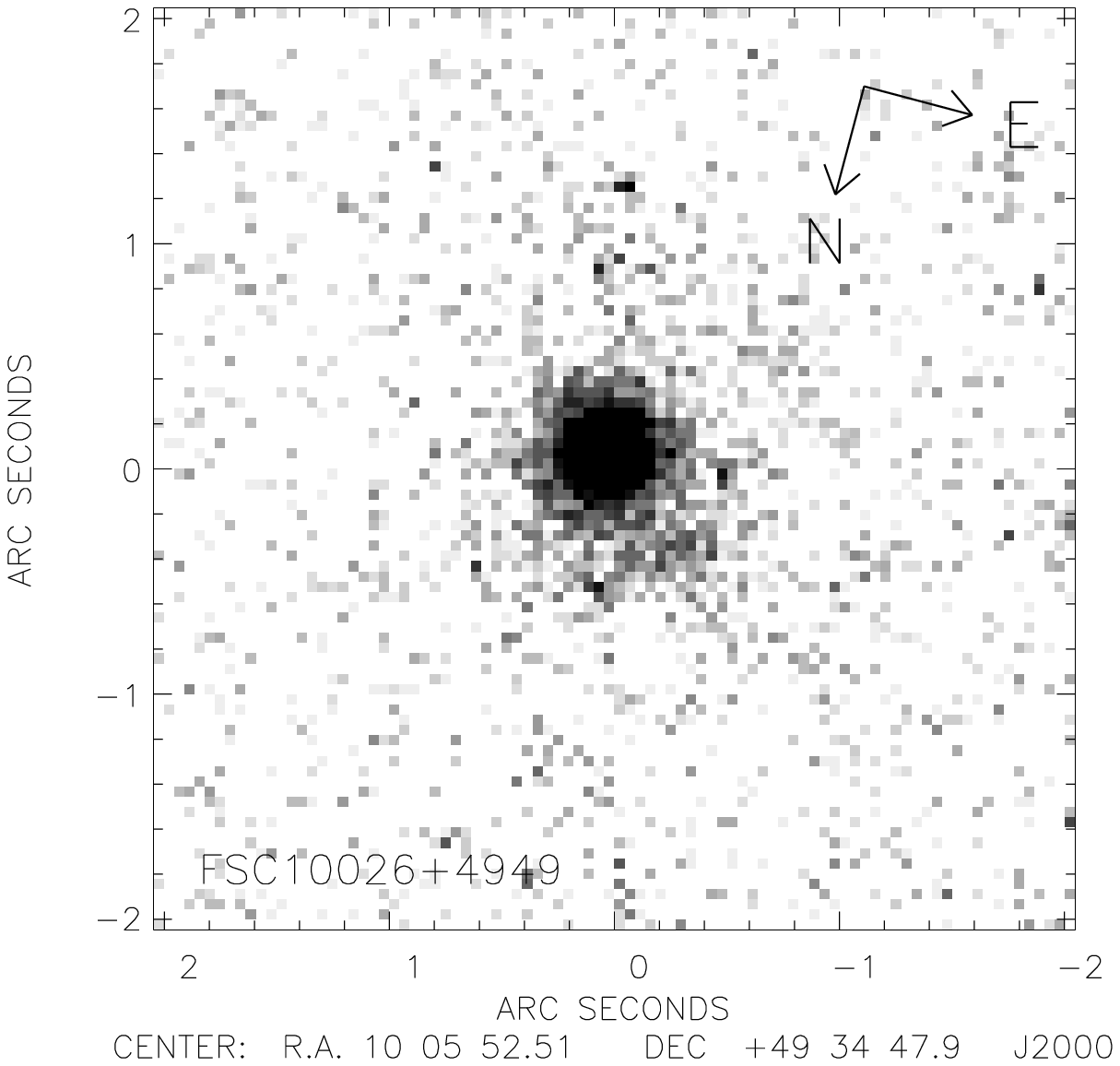,width=55mm}
\epsfig{figure=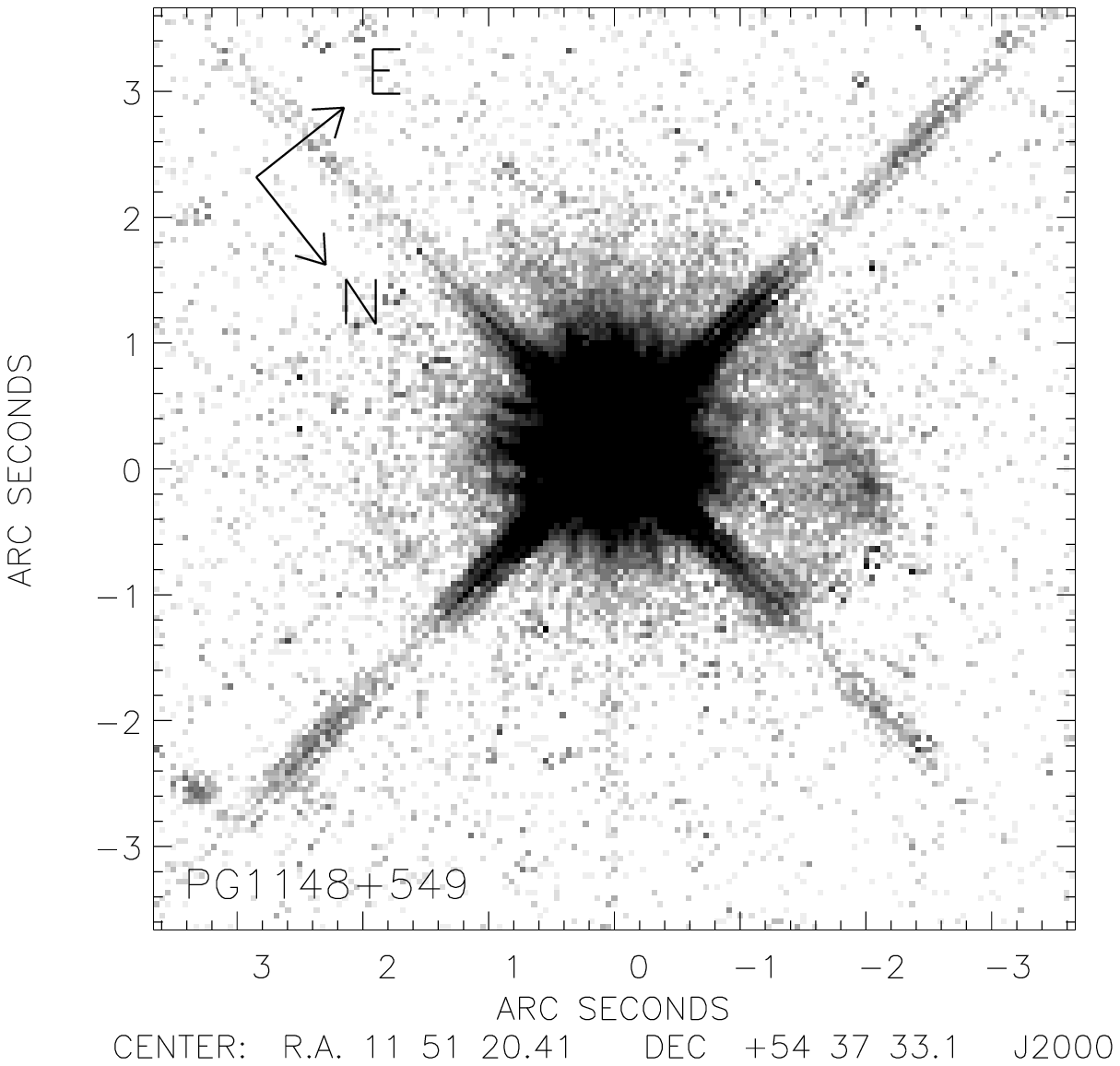,width=55mm}
\end{minipage}
\begin{minipage}{170mm}
\epsfig{figure=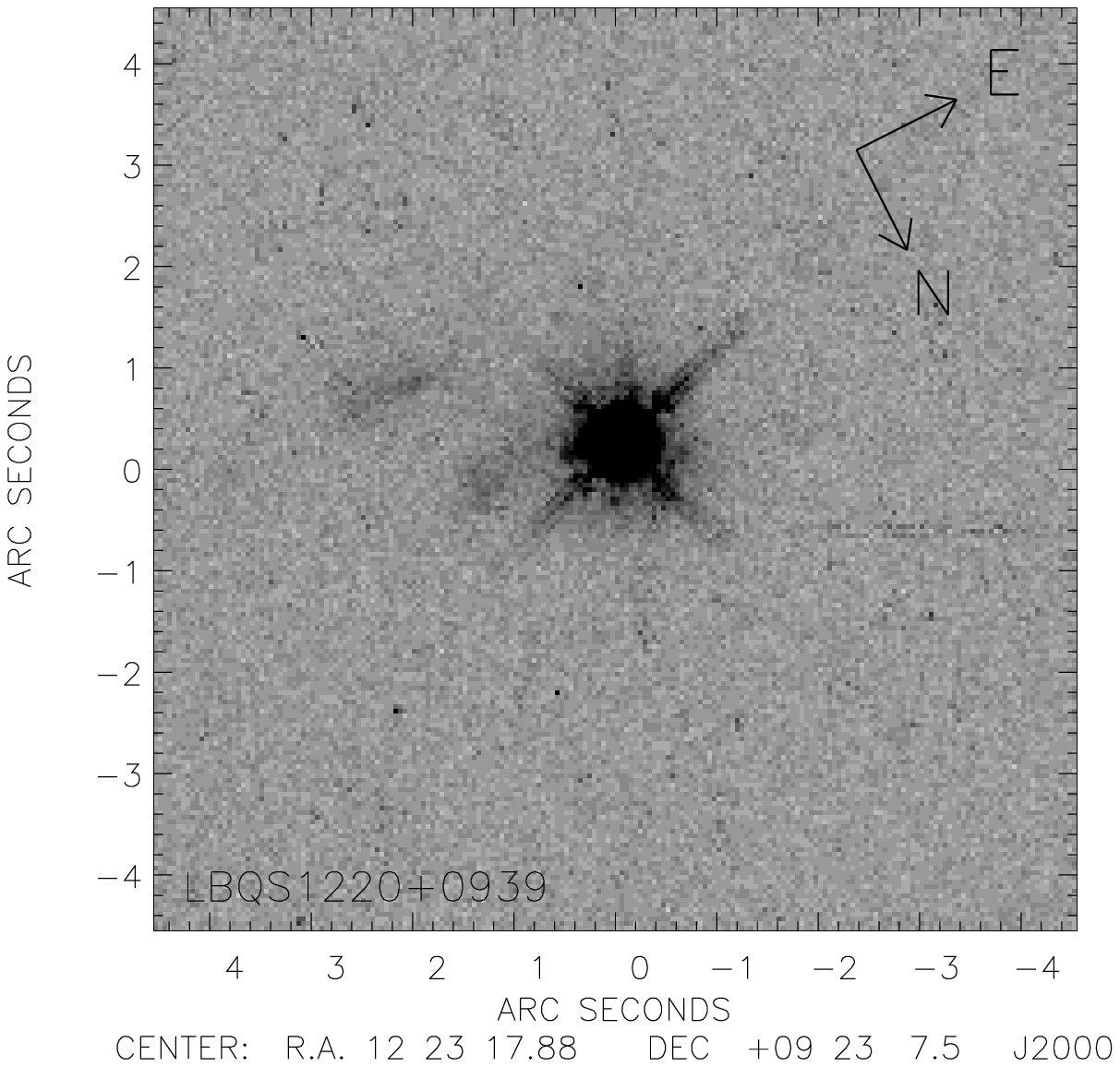,width=55mm}
\epsfig{figure=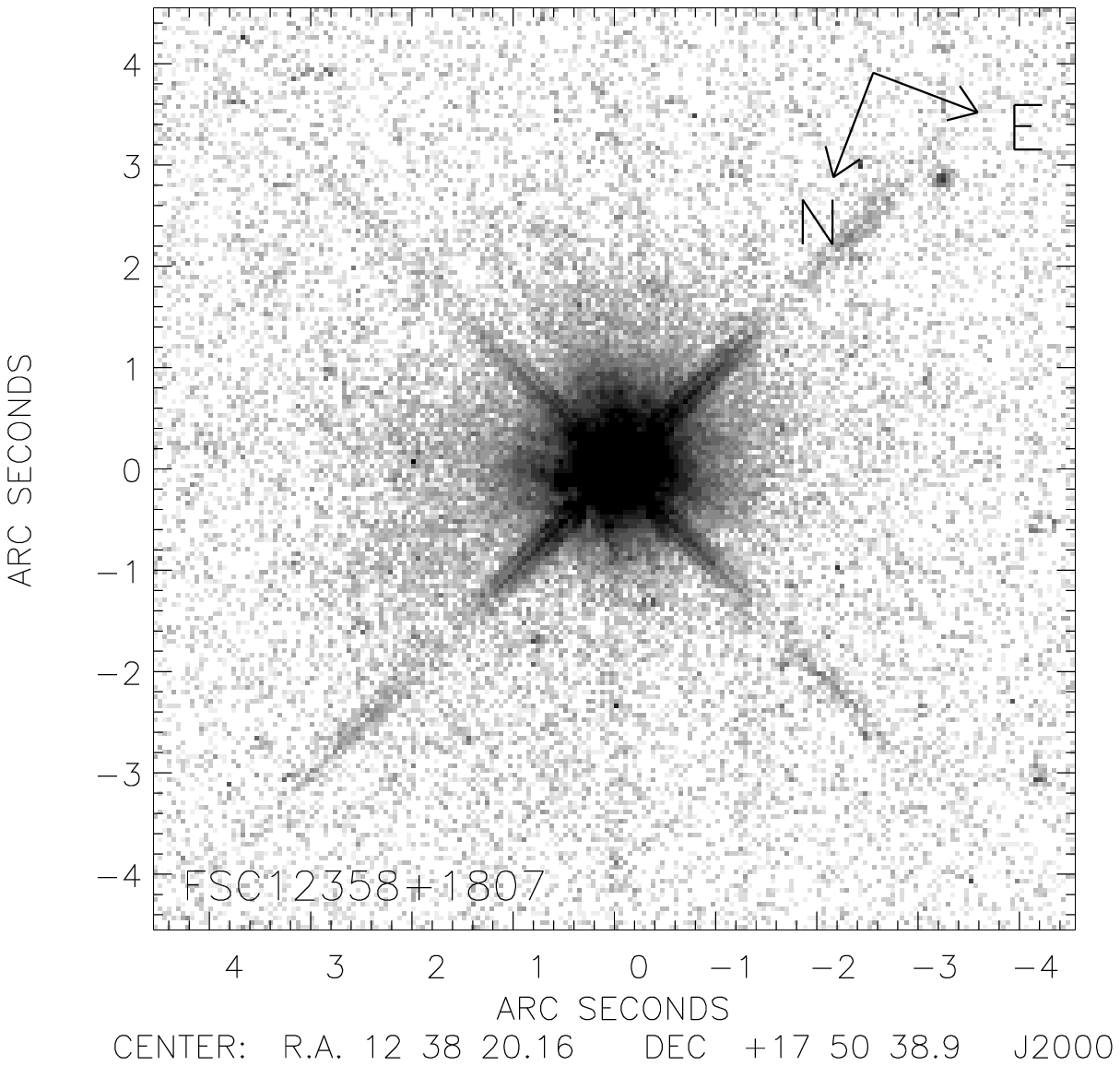,width=55mm}
\epsfig{figure=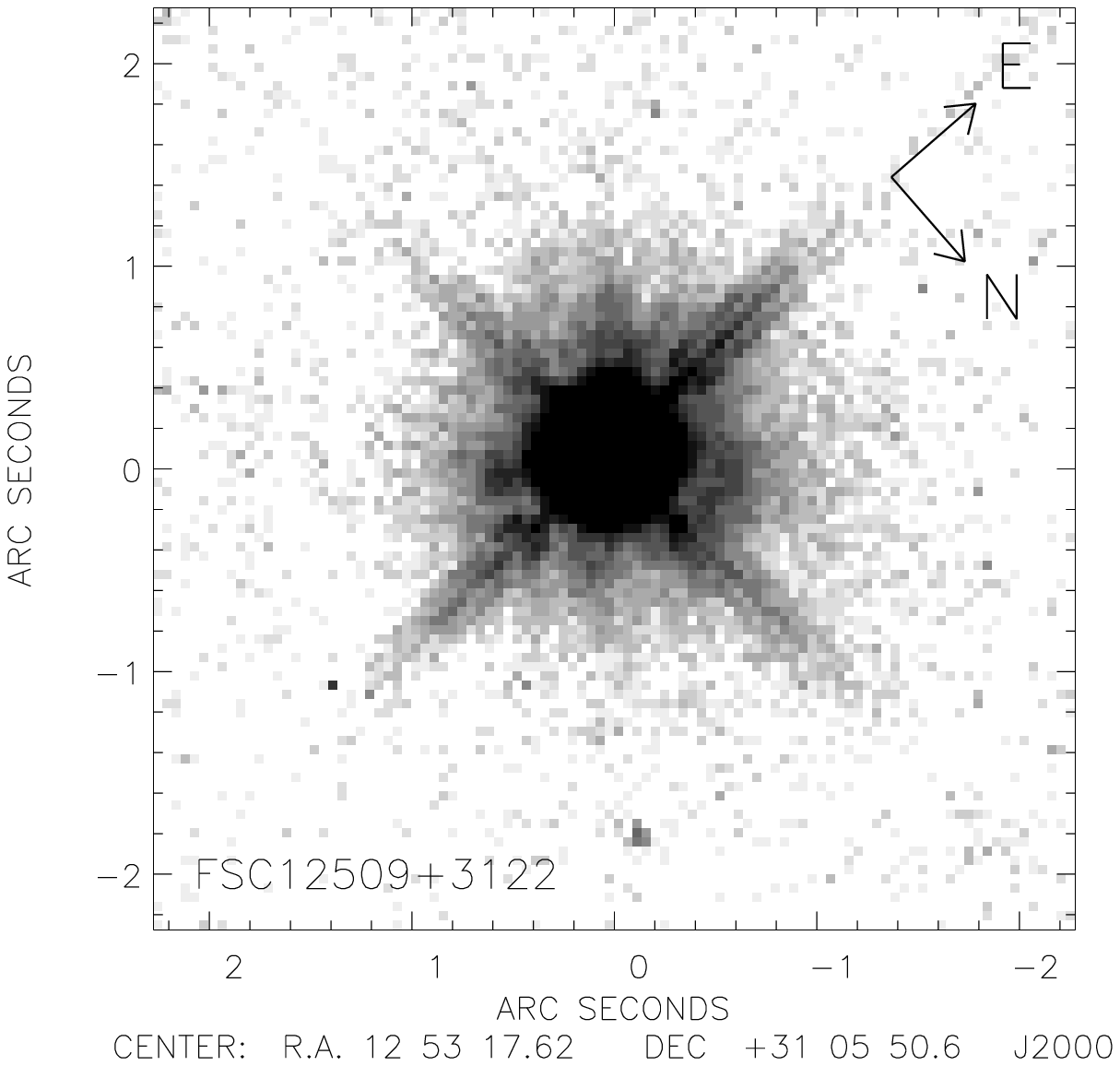,width=55mm}
\end{minipage}
\begin{minipage}{170mm}
\epsfig{figure=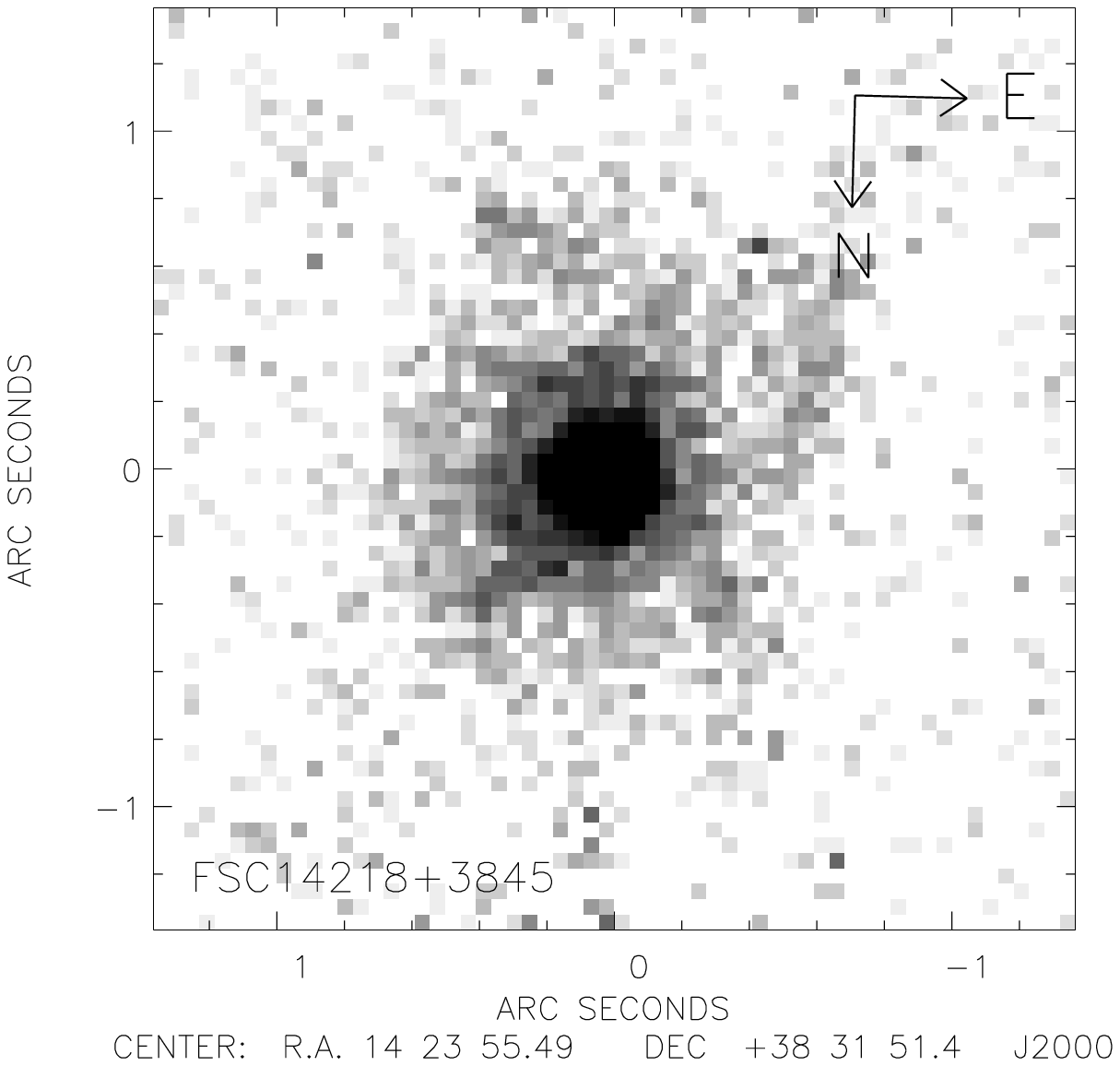,width=55mm}
\epsfig{figure=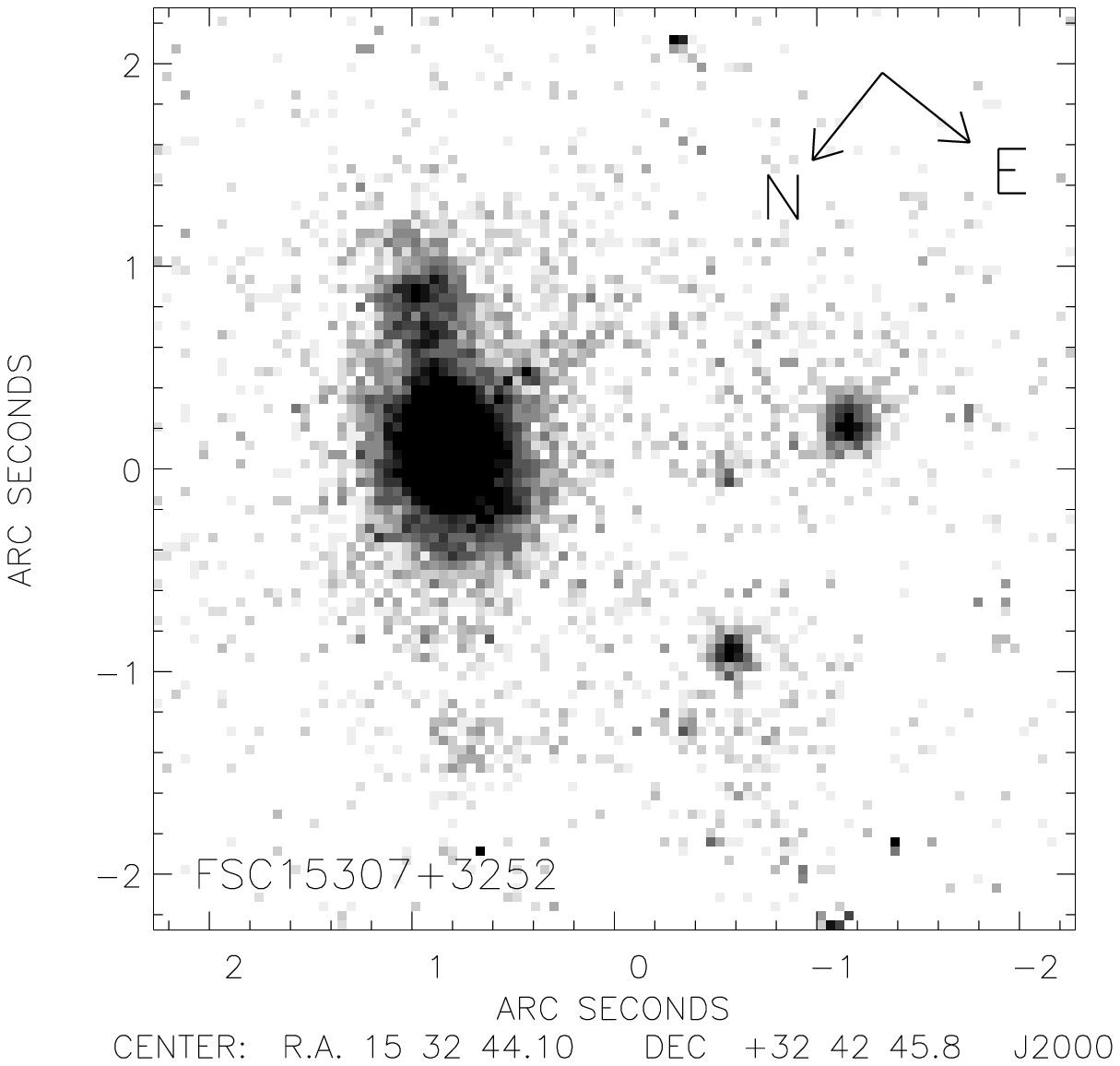,width=55mm}
\epsfig{figure=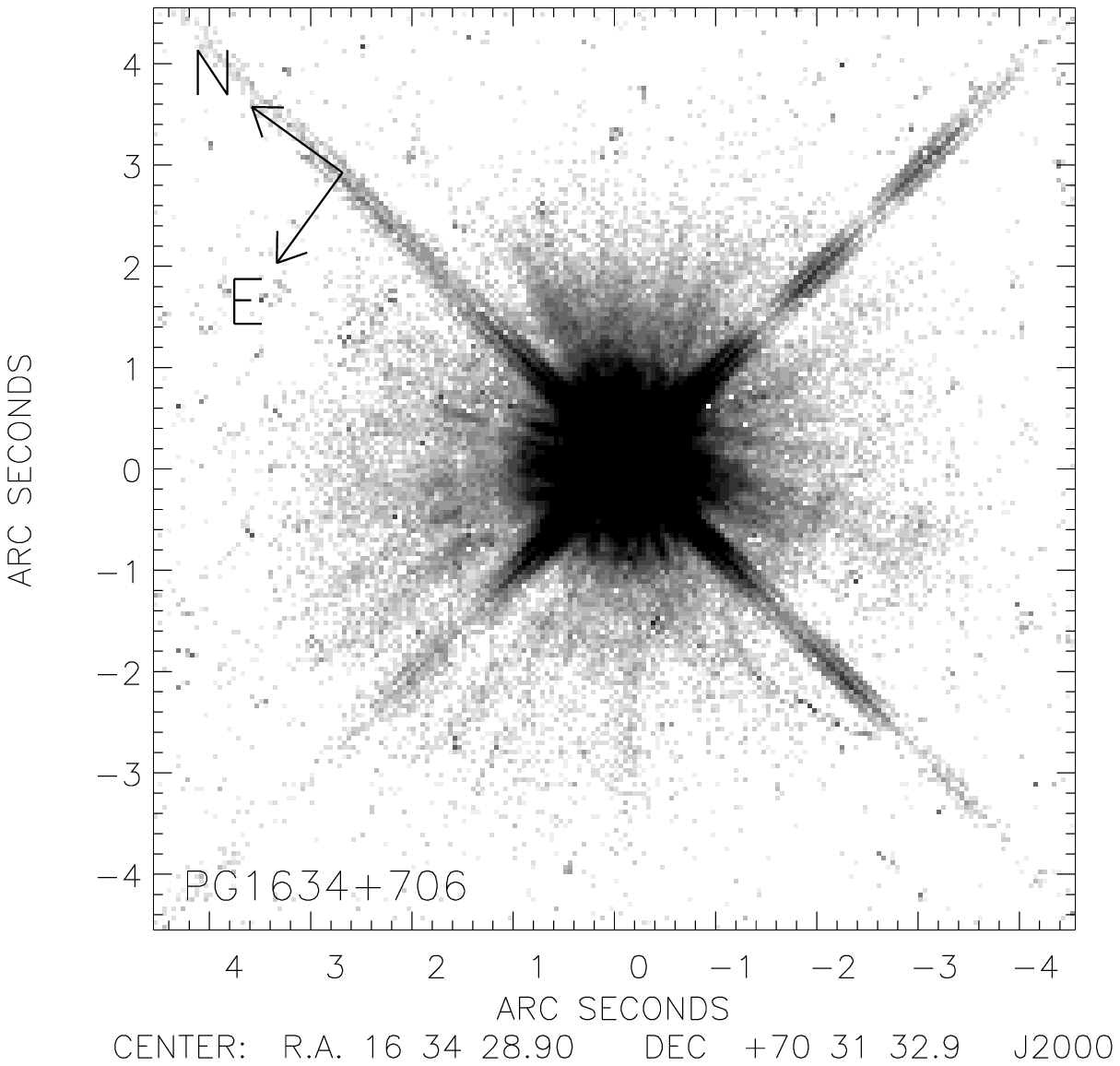,width=55mm}
\end{minipage}
\caption{{\em HST} F814W images of the 9 HLIRGs, showing the details of each source and scaled to show any interesting features.
 \label{hlirgs_im_src}}
\end{figure*}

\begin{figure*}
\begin{minipage}{170mm}
\epsfig{figure=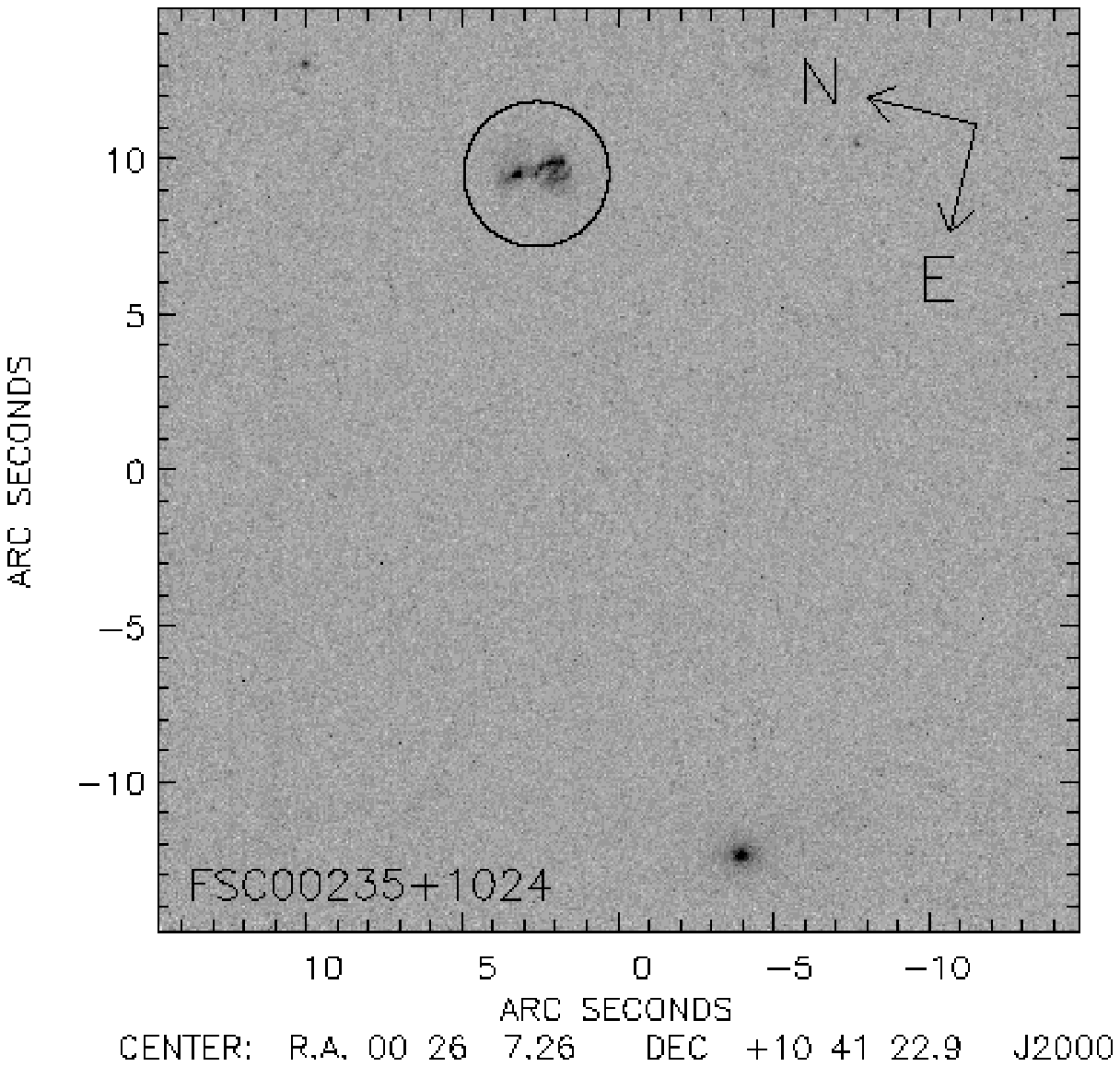,width=55mm}
\epsfig{figure=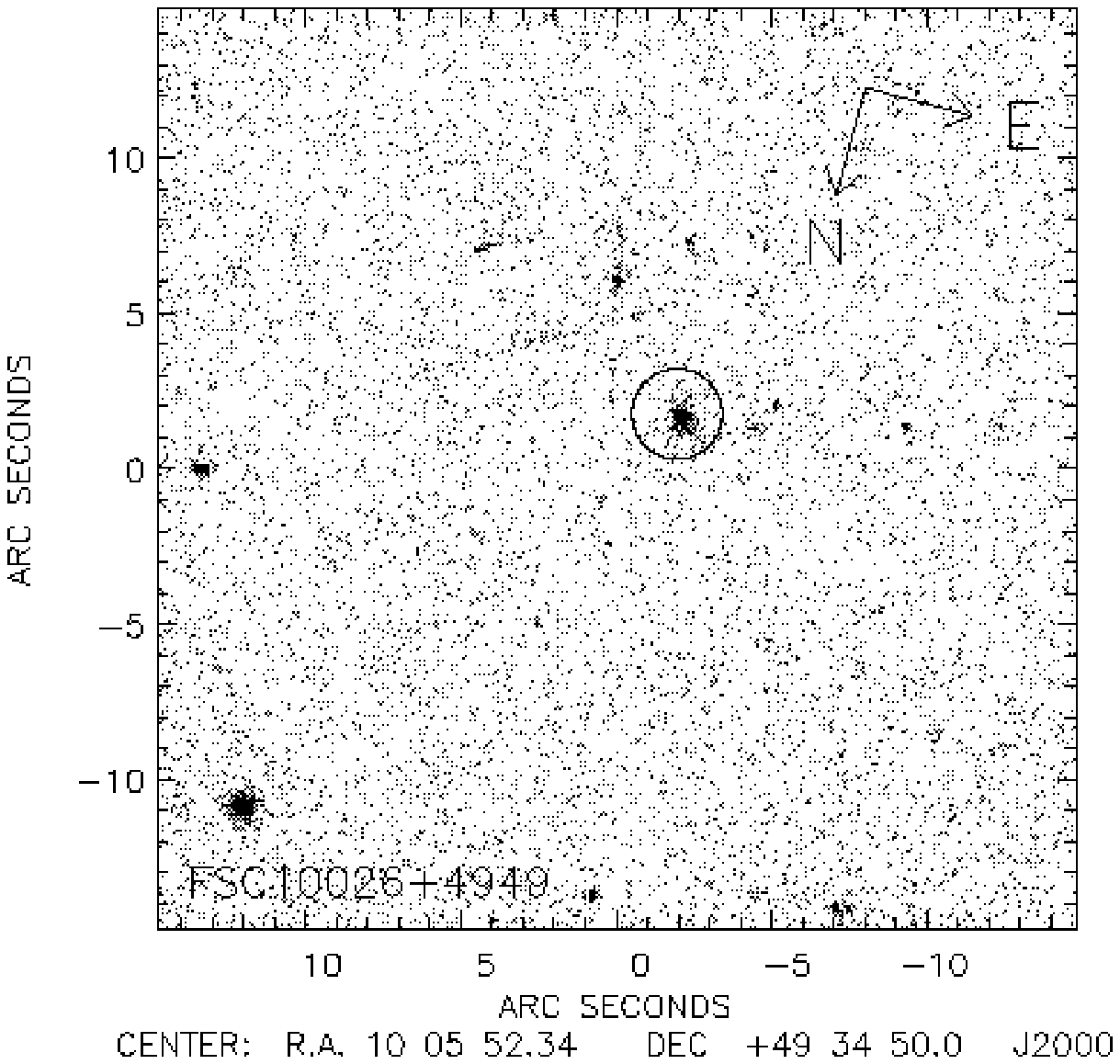,width=55mm}
\epsfig{figure=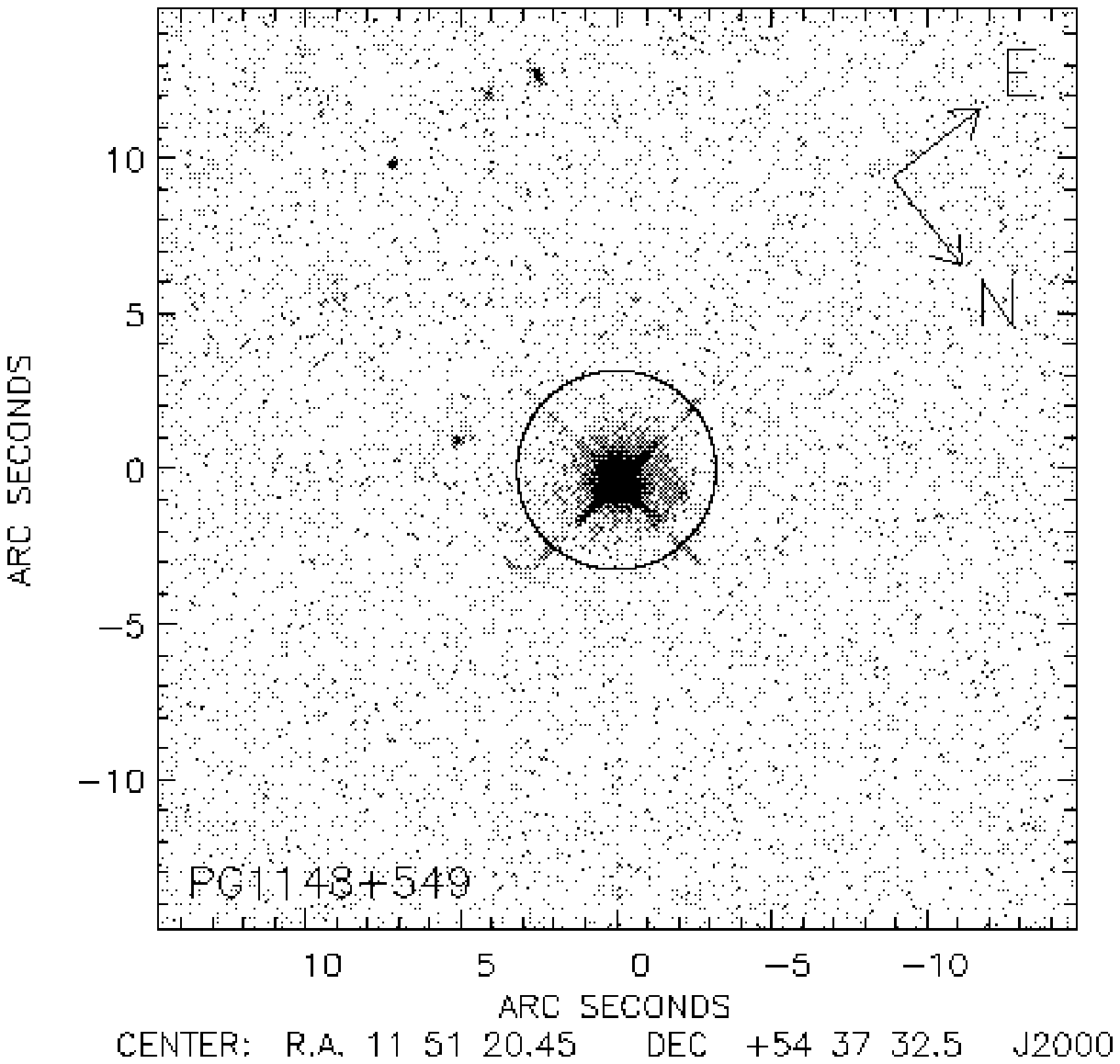,width=55mm}
\end{minipage}
\begin{minipage}{170mm}
\epsfig{figure=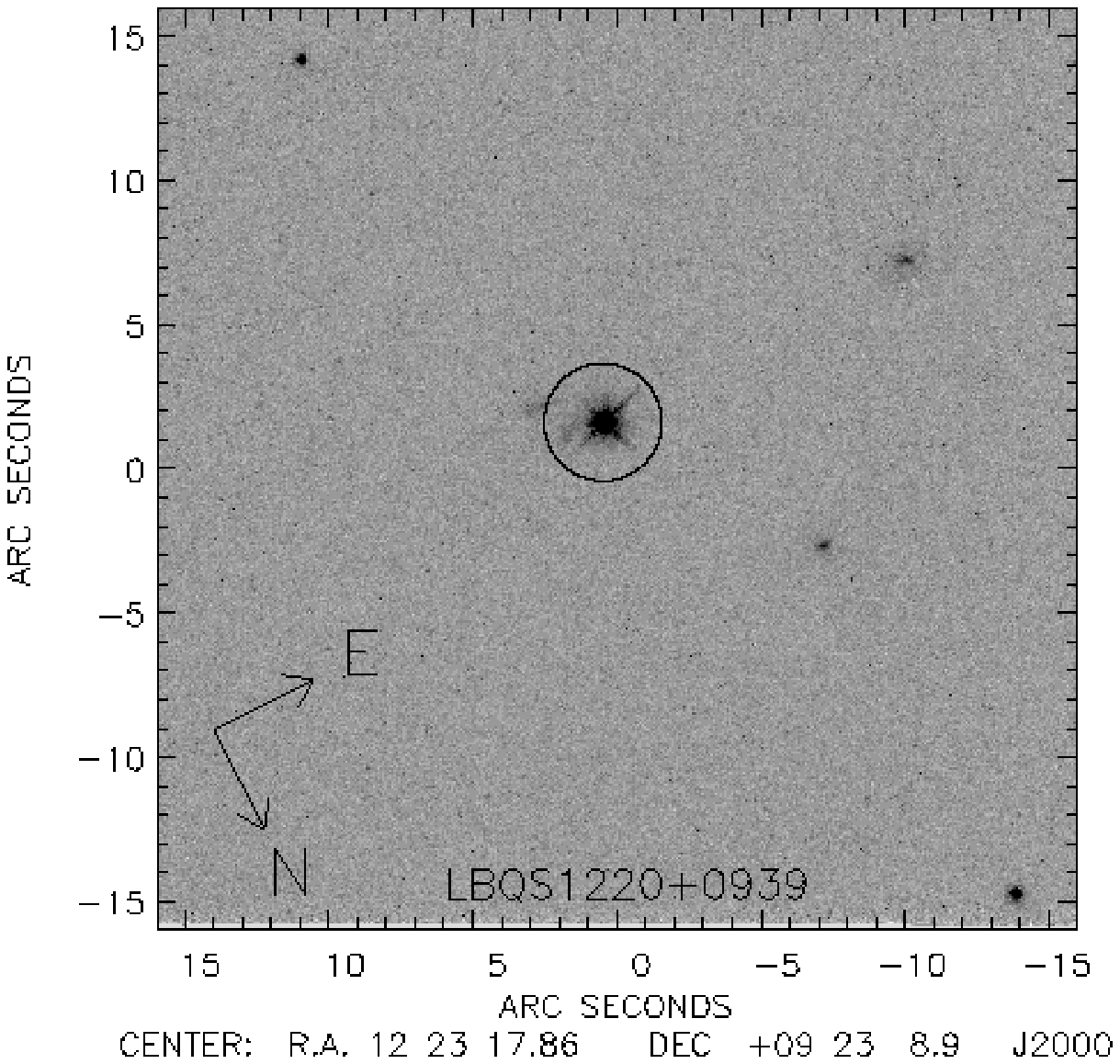,width=55mm}
\epsfig{figure=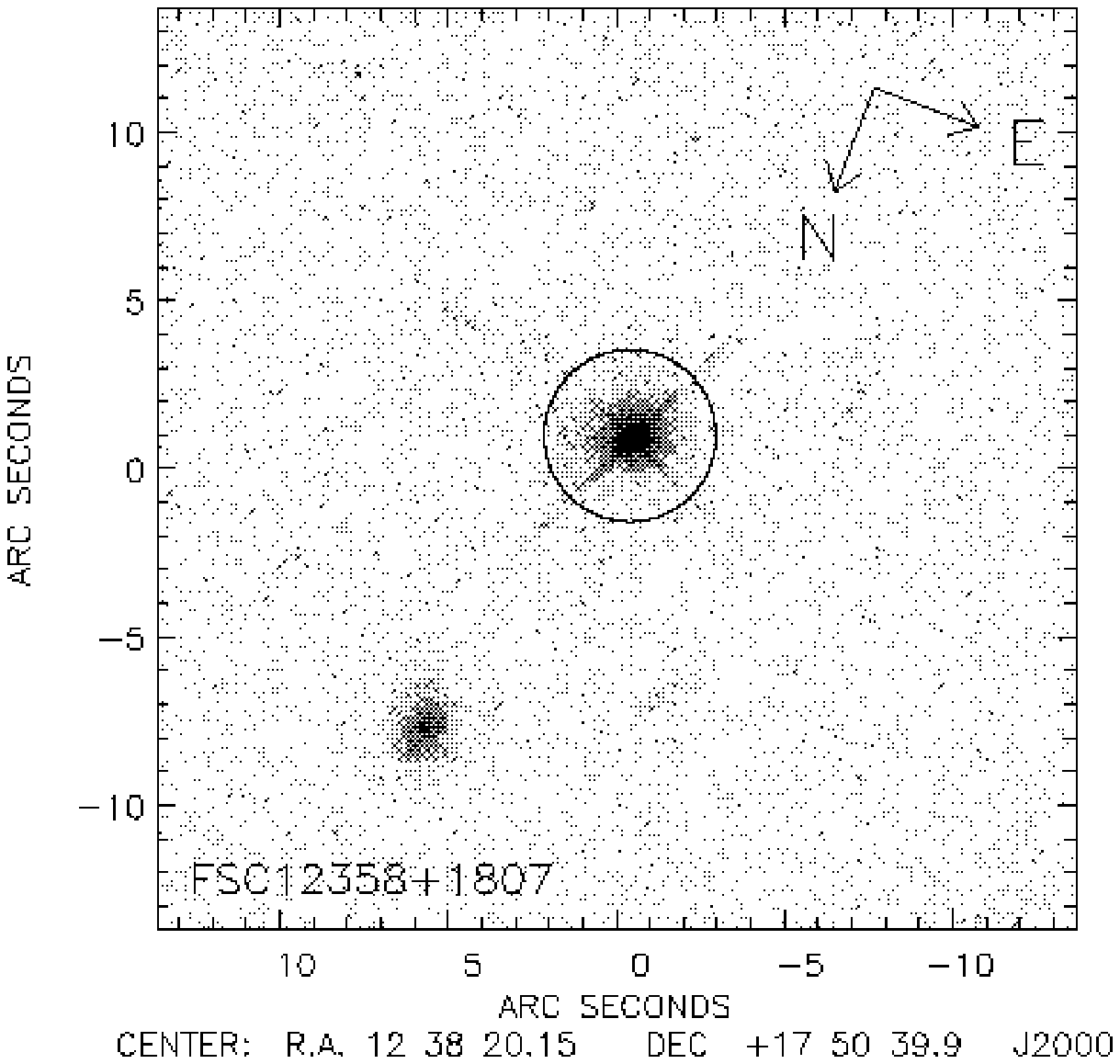,width=55mm}
\epsfig{figure=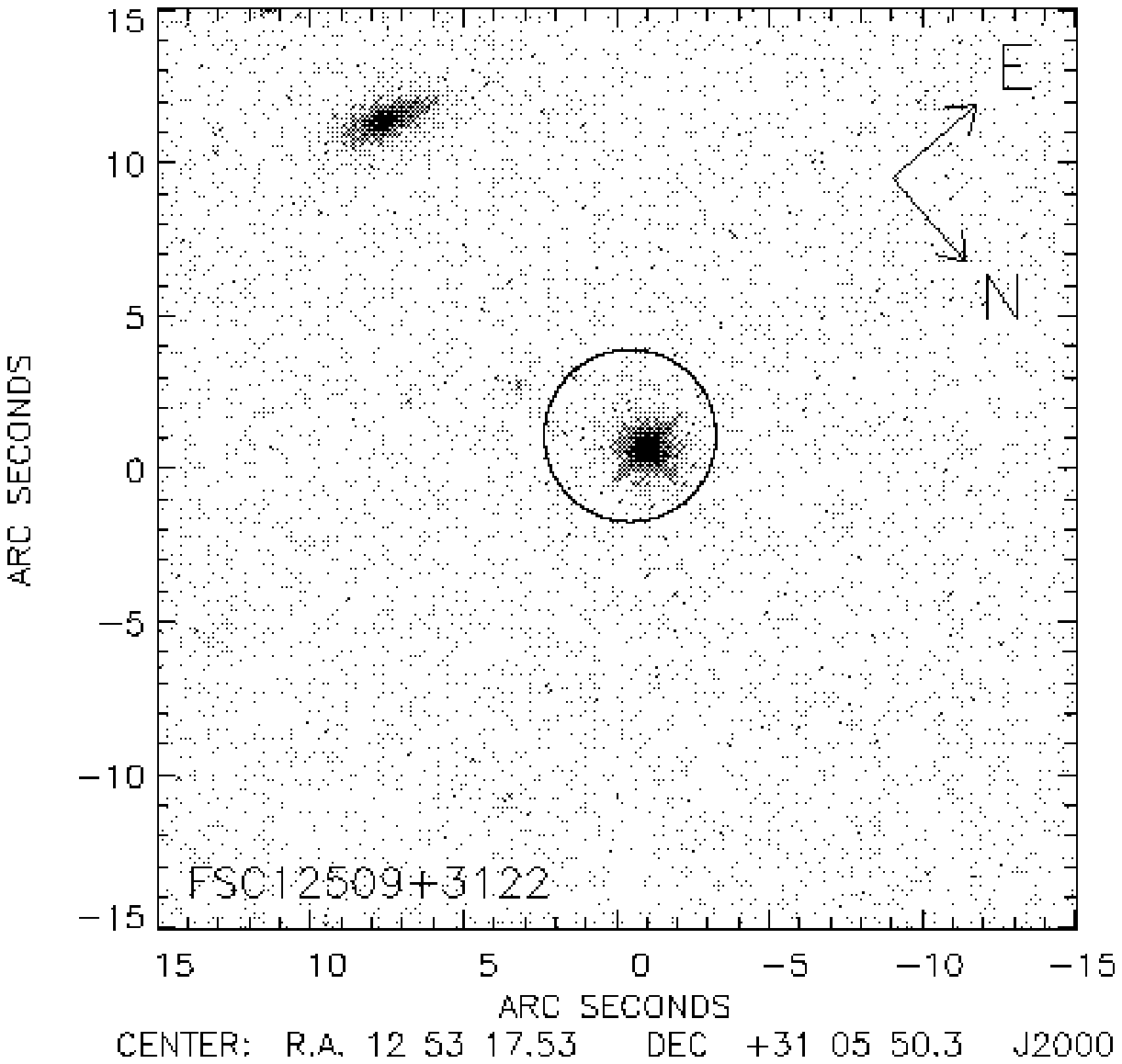,width=55mm}
\end{minipage}
\begin{minipage}{170mm}
\epsfig{figure=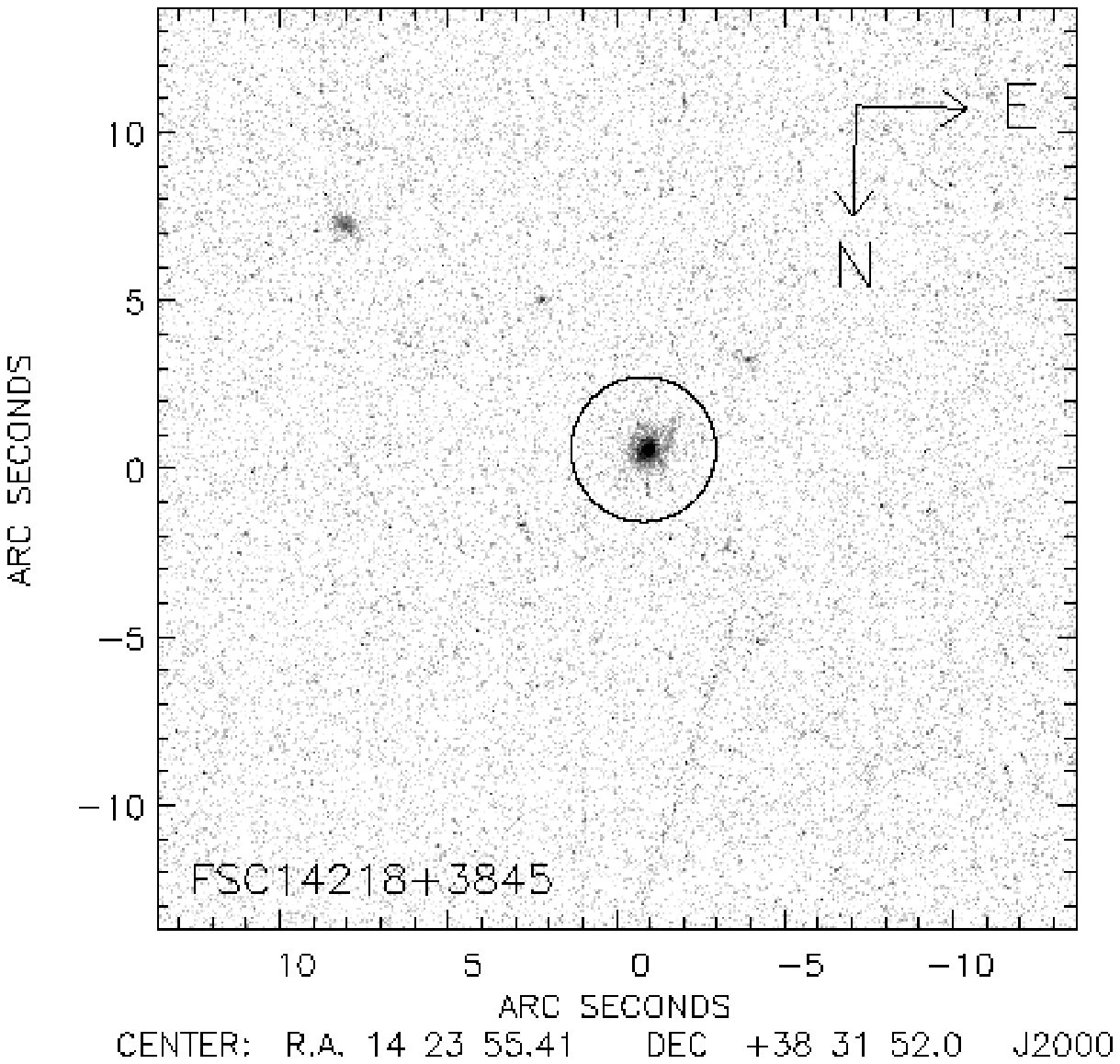,width=55mm}
\epsfig{figure=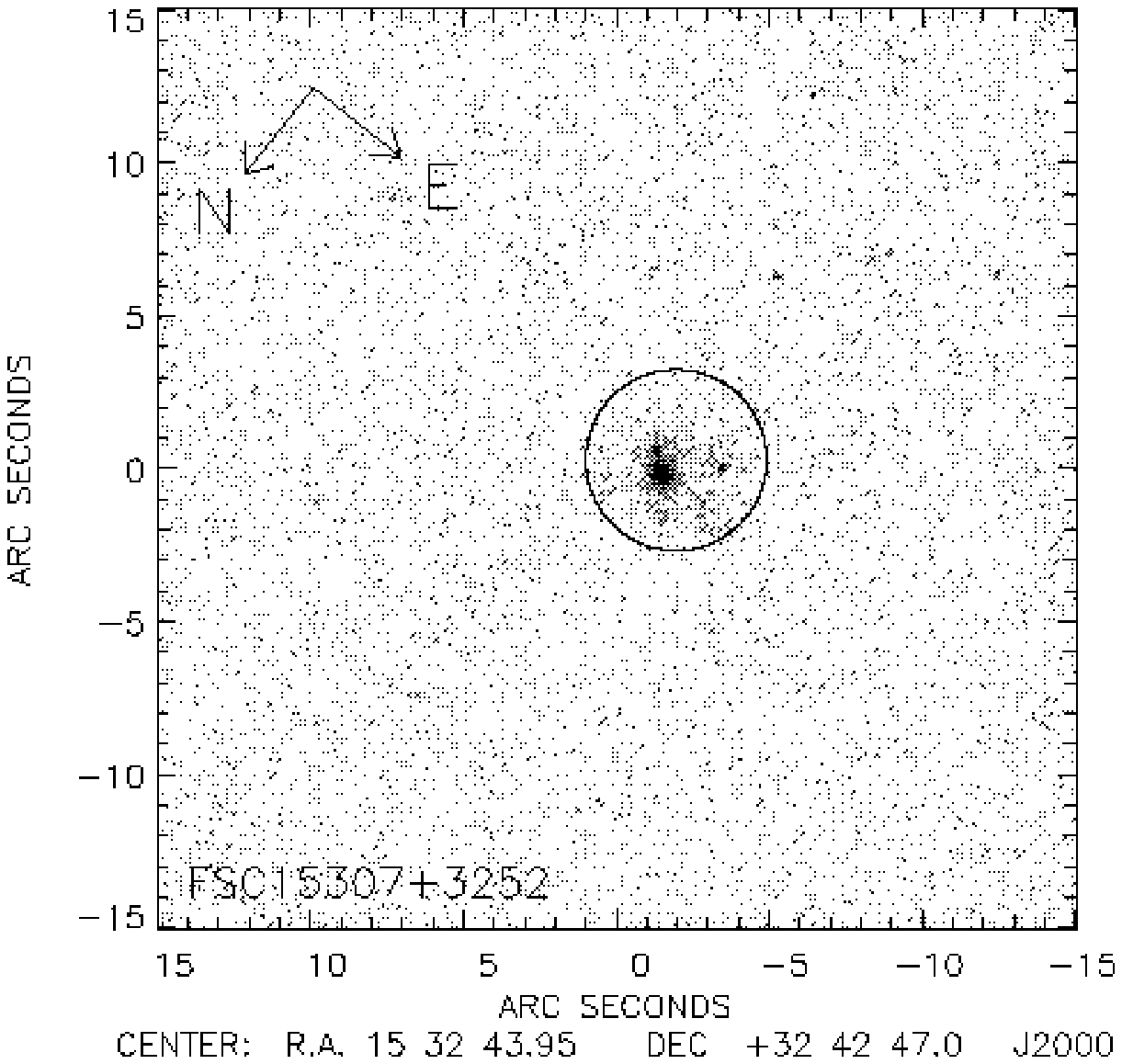,width=55mm}
\epsfig{figure=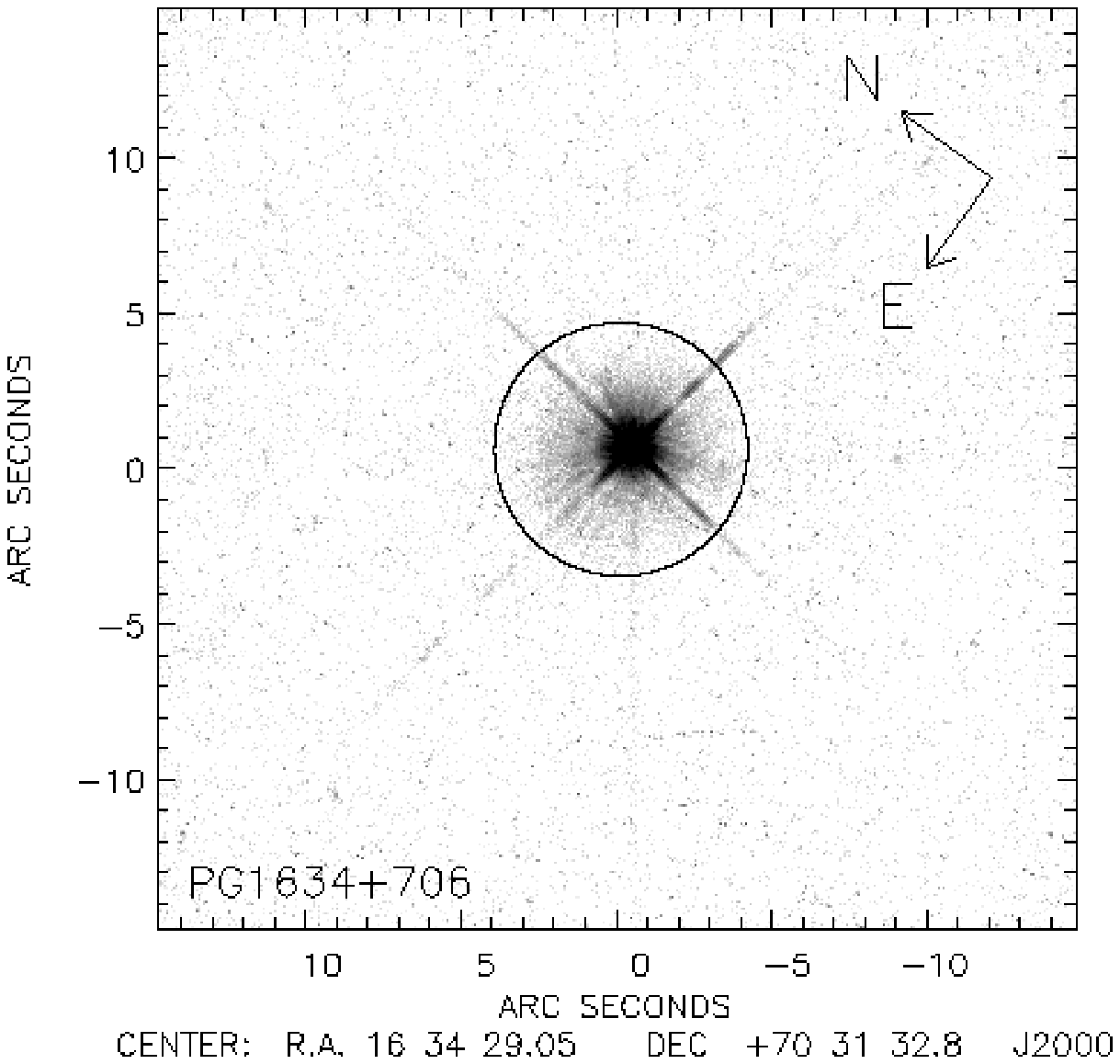,width=55mm}
\end{minipage}
\caption{{\em HST} F814W band images of the 9 Hyperluminous galaxies, showing the sources (circled) and 
immediate environments. \label{hlirg_im_env}}
\end{figure*}

\subsection{The Point Spread Function}
Host galaxies can only be resolved in QSOs if light from the central bright source, and from the host galaxy, can be separated. 
The most convenient way to accomplish this is by fitting a point source template to the QSO; the template being established from 
a set of observations of stellar Point Spread Functions (PSFs) that are a 
good match in filter, colour and detector position to the targets. For the objects in this sample no additional observations were 
made of nearby stars, and no suitable observed PSFs were available in the {\em HST} PSF Archive. Synthetic PSFs were therefore 
generated using the TinyTIM v5.0 software \cite{kr}, recently updated to include field dependent effects such as astigmatism, coma, 
and focus. 

Ten times oversampled PSFs were generated. These PSFs were then shifted by non-integer pixel distances, rebinned to normal 
resolution and convolved with the PC pixel scattering function. The agreement between these synthetic PSFs and observed (stellar) 
PSFs is generally excellent within radii of about 2\arcsec. Beyond this radius scattered light in the WFPC2 optics can have a 
significant effect, an effect that is not modelled by the TinyTIM software. The effect of any systematic uncertainties in the 
PSF have also not been examined, such effects are however likely to be extremely small.

The resulting PSFs were then centroided  and normalized to the image via a reduced $\chi^{2}$ fit. It was found that all the 
images were saturated in the central regions to varying extents, mostly confined to the central $3\times3$ pixels and never beyond the 
central $5\times5$ pixels. In all cases the centralmost possible pixels were used, avoiding those that were saturated or that lay in the 
diffraction spikes. This method allows the image and PSF to be registered to within $\sim$0.2 pixels (estimated by visual 
inspection), and also gives a starting value for the PSF normalisation.

\subsection{Profile Fitting}
In order to determine the morphology of the host galaxies in the sample, both de Vaucoleurs and exponential disk profiles were 
fitted to the surface brightness profiles of the host galaxies. The de Vaucoleurs profile is of the form:

\begin{equation}
I(r) = I_{0}e^{-7.67[(r/r_{e})^{1/4} - 1]}
\end{equation}

\noindent and the exponential disk profile is of the form:

\begin{equation}
I(r) = I_{0}e^{-(r/h)}
\end{equation}

One dimensional surface brightness profiles of the sources were extracted by fitting annuli spaced at 1 pixel intervals, fixing the 
ellipticity (at $\epsilon = 0.1$), semimajor axis and position angle. The PSF profile together with either a radial de Vaucoleurs or 
disk profile were then fitted to the source surface brightness profile, treating the PSF normalisation as a third free parameter. 
This helps to minimise as far as possible the systematic errors introduced by subtracting off a `best guess' PSF and fitting a galaxy 
profile to the remaining light distribution. Fitting was carried out using Levenberg-Marquardt least-squares minimization in order to 
robustly derive the best fit parameters. Two dimensional isophote fitting, using the IRAF task E{\sevensize LLIPSE}, was also used to 
investigate the ellipticities and position angles for each host by iteratively fitting isophotes of constant surface brightness.

In order to quantify whether a host galaxy was detected or not we imposed the condition that a PSF + galaxy to QSO fit must be better 
than a pure PSF fit to at least 95\% confidence, computed via an {\em F} test. We excluded pixels within the central 0.1\arcsec\ 
as pixels within this radius tend to be highly undersampled and in some cases saturated. We also excluded all pixels beyond 2\arcsec\ 
because of light scattered within the WFPC2 optics.

\section{Results}

Coordinates, redshifts and measured F814W magnitudes for the sources can be found in Table \ref{hstobs}, together with 
the optical morphologies, IR luminosities and any previously cited best fit IR SED models. Table \ref{hlirgs_hosts} summarises 
the host galaxy properties, and includes relative/absolute magnitudes, {\em k} corrections and host galaxy properties derived from 
profile fitting. In the sections where host galaxies are discussed, this refers to those objects listed in Table 2, thus including 
three sources which do not possess a central bright source. Discussion of PSF subtraction and/or host galaxy detection limits are 
not applicable to these objects.

\subsection{Morphologies and Lensing}
The {\em HST} images showing the sources and immediate environments can be found in Figure \ref{hlirg_im_env}. Detailed images of 
each object can be found in Figure \ref{hlirgs_im_src}.
 
The {\em HST} images show that the sample contains an interesting mixture of morphologies. Two of the sources, F00235+1024 and 
F15307+3252 show clear signs of strong interaction, features include obvious morphological disturbance, tails, and compact bright 
`knots'. One further source, F10026+4949 is small, shows no bright central source and slightly weaker signs of interaction. One 
object, F14218+3845, is a QSO with a relatively underluminous point source. The five remaining sources are all optically luminous 
QSOs. Although some of the QSOs show some very faint inhomogeneities, none of them show any definite signs of recent or ongoing 
interaction. 

None of the objects in the sample show any evidence for gravitational lensing. We investigated the possibility of lensing for all 
sources using the {\em HST} gravitational lens simulator constructed by Kavan Ratnatunga as part of the HST Medium Deep Survey 
\cite{gri,rat}. For non-trivial magnification arcs of length comparable with the 
Einstein radius should be observed, so to first order we can set limits on the Einstein radius of 0.3\arcsec\ ($\sim$6 PC pixels). 
If any lensing is therefore present it would have to be by low mass galaxies and so would make no appreciable difference to the 
derived optical and IR fluxes, unless the background source happened to be aligned with a caustic. The lack of multiple images 
around any of the QSOs limits any possible lensing magnification to less than a factor of 2 in these objects, assuming an isothermal 
sphere model. If no other signs of lensing are apparent at {\em HST} resolution however, such as tangential stretching, then it is 
extremely unlikely that lensing significantly contributes to the luminosity of any of the objects in the sample. Those sources whose morphologies 
are, on first impressions, plausibly consistent with some form of lensing scenario are discussed further in \S 4.5. 

For two of the three sources not dominated by a bright point source, F10026+4949 and F15307+3252, we were able to fit theoretical galaxy 
surface brightness profiles. For both these sources the galaxies were found to be extremely luminous ellipticals, the properties of which can be 
found in Table \ref{hlirgs_hosts}. The morphology of F00235+1024 is so disturbed that no type of galaxy profile can be fitted to 
the source surface brightness profile.  

For the two PG quasars Schmidt \& Green \shortcite{sc} have measured fluxes at 440nm. Their results, compared to the measured 
F814W fluxes given in Table \ref{hstobs}, suggest that the optical continuum for both these sources is extremely flat, with $\alpha \sim 0$.

\subsection{Infrared Properties}
Table \ref{hstobs} lists the IR properties of the sample. Three of the sources (F00235+1024, F14218+3845 \& F15307+3252) have been 
previously observed by the Infrared Space Observatory ({\em ISO}) as part of a larger sample, and their Spectral Energy Distributions modelled 
using a suite of AGN and starburst models \cite{ve}. The IR emission from PG1634+706 has been previously modelled by Haas et al. 
\shortcite{haa} using a series of greybody functions. The IR emission from a further three sources (F10026+4949, F12509+3122 \& PG1148+549) has 
been previously examined by Rowan-Robinson \shortcite{rr2}. For the final two sources in the sample (LBQS1220+0939 \& F12358+1807) there 
exists insufficient photometry to model the power source behind the IR SED. Some constraints on the IR emission from F10026+4949 \& F12509+3122
can also be drawn using the {\em IRAS} colours. A `warm' infrared colour (i.e when the IRAS $25\mu$m to IRAS $60\mu$m flux ratio has a value of 
$f_{25}/f_{60}\geq0.2$) generally indicates the presence of an AGN \cite{deg}. Both F10026+4949 ($f_{25}/f_{60}=0.665$) and F12509+3122 
($f_{25}/f_{60}=0.472$) possess `warm' colours, and have been previously cited as containing an AGN by Rowan-Robinson \shortcite{rr2}.
For F12358+1807 and LBQS1220+0939 the {\em IRAS} fluxes and upper limits do not allow us to distinguish between `warm' and `cool' IR colours, 
hence the power source is unconstrained.

For two of the sources in the sample (PG1148+549 \& PG1634+706) we present new fits to the IR SED. To study the IR emission from these 
two sources, we have fitted the SEDs using combinations of the following two models:

\noindent (1) The inclined AGN dust torus model (Andreas Efstathiou, priv. comm.) is based on previous AGN models \cite{ehy,err}.
The model contains 15 different line of sight orientations, with a torus opening angle of $60$\degr\ and a UV optical depth of
$\tau_{UV}=1000$.

\noindent (2) The pure starburst model \cite{ef1} is based on an ensemble of HII regions including an evolving population of 
young stars. It includes a simple model for the evolution of HII regions, and a dust grain model incorporating Polycyclic 
Aromatic Hydrocarbons \cite{sie}. The set of models used have a star formation rate exponentially decaying at a fixed rate 
($\tau=20$Myr), but with varying initial UV optical depth (4 discrete values) and starburst lifetime (7 discrete values). 

For PG1148+549 we have compiled additional IR photometry to that used in Rowan-Robinson \shortcite{rr2}. We compiled the SED 
using {\em IRAS} fluxes, together with photometry from EUV spectra \cite{tri}. The {\em I} band flux measured in this paper is 
in good agreement with this EUVE photometry. PG1634+706, although previously modelled in the IR using a suite 
of greybodies, has not been previously modelled using the latest generation of radiative transfer codes. For PG1634+706 the FIR emission 
has been well sampled by {\em ISO} \cite{haa}, and we also include the {\em IRAS} fluxes. For both sources we also use the {\em I} band 
fluxes taken from the {\em HST} images in this paper.

\begin{figure*}
\begin{minipage}{180mm}
\rotatebox{90}{
\scalebox{0.35}{
\includegraphics*[0,0][504,710]{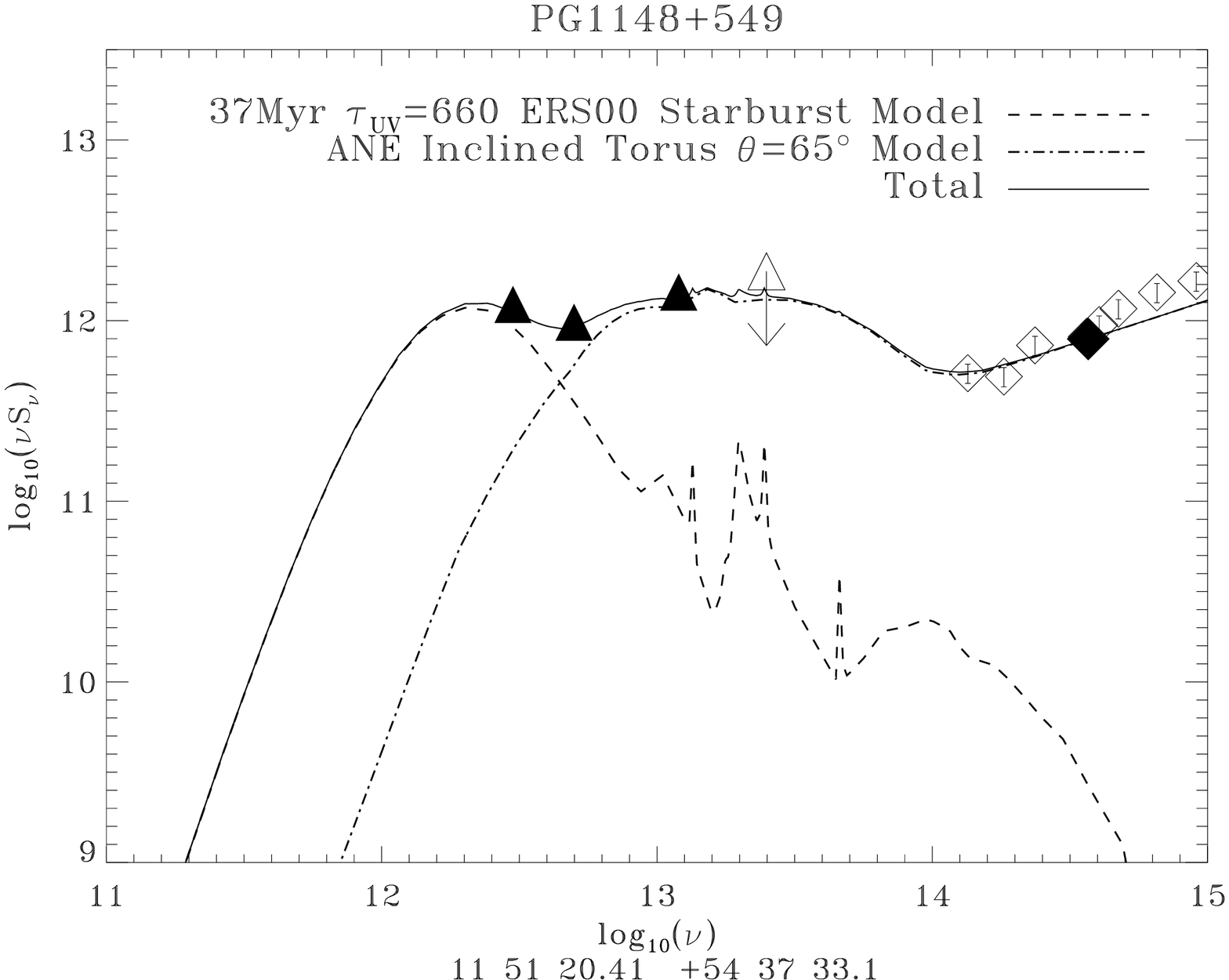}
}}
\rotatebox{90}{
\scalebox{0.35}{
\includegraphics*[0,0][504,710]{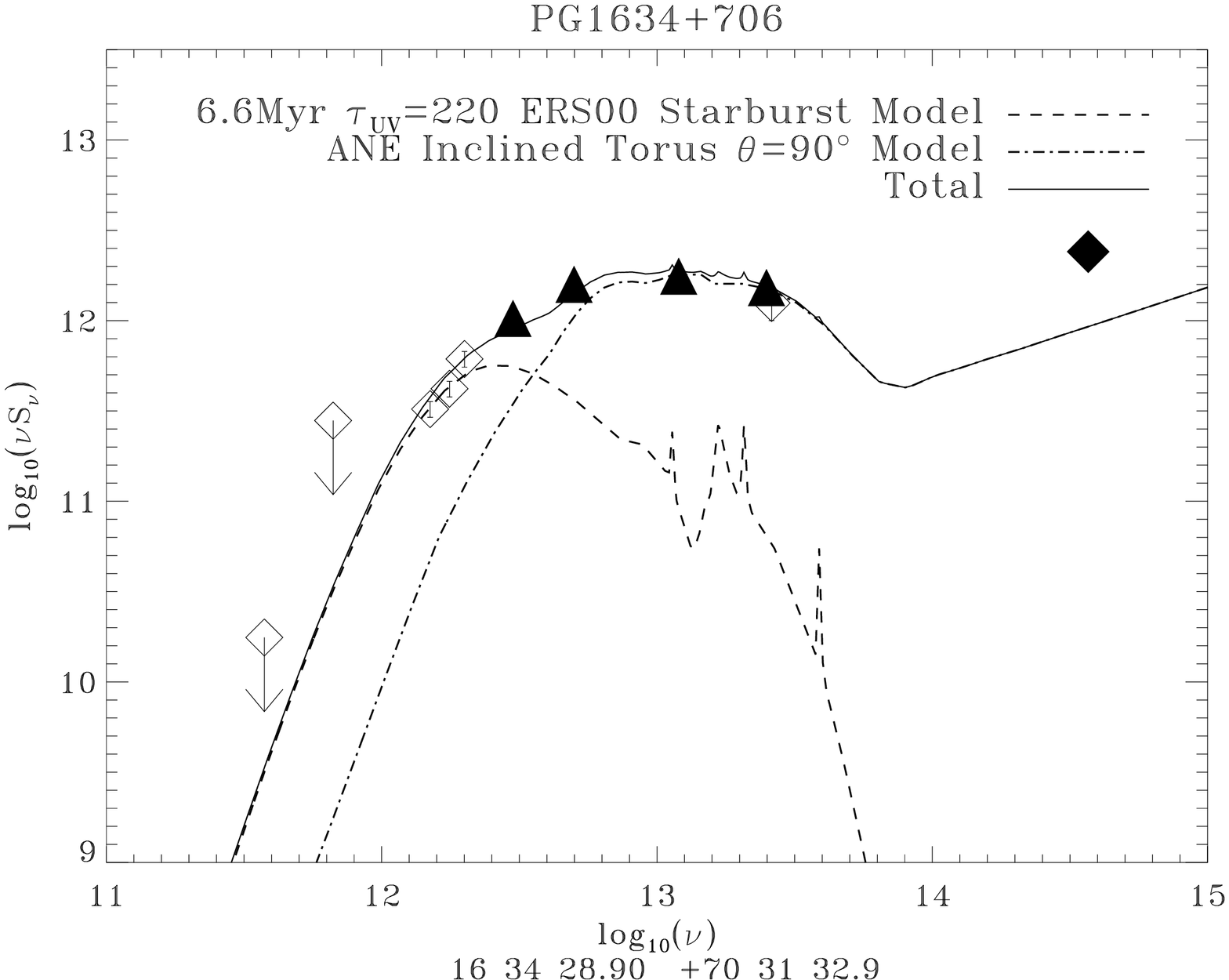}
}}
\end{minipage}
\caption{IR Spectral Energy Distributions for PG1148+549 \& PG1634+706. 
The filled triangles are {\em IRAS} fluxes, and the unfilled triangles with an arrow below them are {\em IRAS} upper limits.
Filled diamonds are the {\em I} band fluxes from this paper. Unfilled diamonds are any additional data taken from the literature. 
For PG1148+549 the additional data is taken from Tripp et al (1994). For PG1634+706 the additional data is 
taken from Haas et al (1998). \label{hlirgs_seds}}
\end{figure*}

To fit the SEDs we have used the two models described above, allowing each model to vary freely in normalization. The starburst 
model is also allowed to vary in the discrete values of initial UV optical depth and starburst age. The torus model 
is also allowed to vary in inclination angle. The two models are then added together, and the reduced $\chi^{2}$ estimator is
calculated for the total of every combination of starburst plus AGN model. The best fit model is then selected using the miniumum 
$\chi^{2}$. The SEDs and best model fits for these two sources can be found in Figure \ref{hlirgs_seds}.

For PG1148+549 the fit to the IR SED is excellent. For PG1634+706 the fit is again very good, 
except that it underpredicts the {\em I} band flux and very slightly overpredicts the FIR emission. 
The small inconsistencies in the far-IR is likely due to having only a 
small number of input models with discrete parameter values, which do not sample the full 
parameter space of either the starburst or AGN models. The underprediction of the {\em I} band 
flux in the SED is most likely also due to this, but could also be accounted for by an unquantified 
contribution from evolved stellar populations which are not included in either model.

\subsection{Companion Sources}
An $L^{*}$ galaxy at the highest redshift in our sample ($z = 1.33$) would have a magnitude of approximately $m_{I} = 22.2$, 
and the faintest galaxies visible (to a $3\sigma$ detection) in any of our observations have a magnitude of $m_{I} \sim 22.5$. 
Companion sources, if present, should certainly be visible for all the sources in the sample. 

Eight of the nine objects in the sample show at least one other object within the PC field of view. Two objects, F12358+1807 and 
F12509+3122 show a single large, bright companion whose magnitude and separation from the source plausibly suggest a cluster environment. 
F12358+1807 also has a number of smaller, dimmer companions. The other six objects all show at least two small, dim objects 
within 15\arcsec. For PG1634+706 there appear to be no other sources visible within the PC field of view. At the redshift of this 
source ($z = 1.33$), this would mean that there are no companions with an optical luminosity comparable to or brighter than $L^{*}$ 
within 0.5Mpc, suggesting that this source is isolated.

\begin{table}
\caption{Host Galaxy Properties \label{hlirgs_hosts}}
\begin{tabular}{@{}ccccccccc}
Galaxy         &$m_{I}$    & $k^{1}$ & $M_{I}^{2}$ & $r_{e}$ (Kpc) & $I_{e}$         \\
F00235+1024  &   -       & -1.18   &-24.7        &     -            &        -        \\
F10026+4949  &   -       & -1.25   &-25.5        & 6.7$\pm$0.8      & 20.6$\pm$0.16   \\
1220+0939    &   19.0    & -1.27   &-25.0        & 87.9$\pm$8.0     & 24.8$\pm$0.15   \\
F12358+1807  &   18.3    & -0.94   &-24.5        & 7.1$\pm$0.29     & 20.8$\pm$0.1    \\
F12509+3122  &   19.1    & -1.34   &-25.2        & 10.3$\pm$0.67    & 21.0$\pm$0.13   \\
F14218+3845  &   20.5    & -1.10   &-24.7        &  -               & -               \\
F15307+3252  &   -       & -1.37   &-26.4        & 12.4$\pm$0.55    & 21.1$\pm$0.79   \\
\end{tabular}

\medskip
F00235+1024, F10026+4949 and F15307+3252 are included here for completeness, in these cases the `host' is 
in fact the source itself. All reliably determined profiles were ellipticals. Although a host for F14218+3845 
was resolved the profile was undetermined. The morphology of F00235+1024 is too disturbed to fit a profile to. 
2$\sigma$ errors are quoted. $^1${\em k} correction for the F814W filter. $^2$Absolute magnitudes including {\em k} 
corrections. 
\end{table}

\subsection{QSO Host Galaxy Properties}
Six objects in our sample are QSOs as determined from the optical morphologies, and host galaxies are clearly resolved in four of 
these six objects. Images of the four host galaxies can be found in Figure \ref{hlirgs_im_host}. 
For a positive host galaxy detection our original imposition of a minimum 95\% confidence limit was found to be easily satisfied, 
all detections achieving at least a 97.5\% confidence with two exceeding 99\% confidence. For the two PG quasars a combination of 
distance and extremely luminous central point source meant that the host was unresolved. The average value of the QSO host galaxy 
magnitude was found to be:

\begin{equation}
\langle M_{I} \rangle = -24.9 \pm0.4 
\end{equation}

\noindent which is comparable in value to the magnitudes of the brightest cluster elliptical galaxies found in the Virgo 
and Coma clusters \cite{ca,jo}. It is also approximately 1.7 magnitudes brighter than the field galaxy magnitude, $M^{*}_{I} 
= -23.2$ \cite{poz,ef}. For comparison the three objects which do not possess a bright central source are, on average, 
about 0.5 magnitudes brighter than the mean QSO host galaxy magnitude, with $\langle M_{I} \rangle = -25.5 \pm0.9$.

For three of the four QSOs in which a host was detected, the residual light profile was sufficiently bright and extended to allow 
a galaxy profile to be reliably fitted. All three hosts were found to be very luminous ellipticals, with large derived scalelengths. 
These values can be found in Table \ref{hlirgs_hosts}. The scalelength for LBQS1220+0939, at $\sim 88$Kpc is extremely large, and 
is comparable to the largest observed ellipticals. Measured ellipticities were found to be very small, generally $0.1 < \epsilon 
< 0.2$, justifying our use of a fixed ellipticity in extracting surface brightness profiles.

\subsection{Individual Sources}

\subsubsection{F00235+1024}
$z = 0.58$. There is no X-ray detection down to a
limit of $L_{x}/L_{bol}=2.3\times10^{-4}$, indicating 
either atypically weak X-ray emission or an obscuring column density of
$N_{H}>10^{23}cm^{-2}$ \cite{wi}. The IR power source is predominantly of 
starburst origin, but with a significant ($37\%$) AGN contribution \cite{ve}.

The {\em HST} image shows an object which is clearly strongly interacting.
There is a compact,  bright region to the north, with 
traces of nebulosity extending further to the north. To the south the
emission is less bright, more extended and shows some small 
plumes. Present throughout this region is a series of very compact,
bright, `knots'. There are two small companion sources; one 
6.9\arcsec\ to the north and one 23\arcsec\ to the east.

\subsubsection{F10026+4949}
$z = 1.12$. The IR power source of this object is best explained as being AGN dominated, but 
with an uncertain starburst contribution \cite{rr2}. The `warm' IR colour ($f_{25}/f_{60}=0.665$) also 
implies AGN dominance. The object is small (less than 1\arcsec\ in diameter), has a very low ellipticity 
($\sim 0.1$) and has no bright central source. There is some 
morphological disturbance in the form of faint nebulosity extending
towards the east. Six small, dim companion sources are 
visible within a 6\arcsec\ radius, two of which lie within 3\arcsec. These
very close companions and asymmetric light 
distribution imply recent or ongoing interactions. The surface brightness
profile is best fitted by an elliptical profile, 
with $M_{I} = -25.5$ and a scalelength of 6.7 Kpc.

\subsubsection{PG1148+549}
$z = 0.97$. This radio quiet QSO is cited \cite{ha} as probably containing
the broad emission line Ne VIII. This would require 
a total column of $N_{H} > 10^{22} cm^{-2}$, and at least 33\% obscuration
of the central regions. The presence of this line 
would mean that gas in the broad emission line regions would be extremely
hot, and that this component would appear as an X-ray 
warm absorber if it lies along the line of sight to the X-ray continuum
source. Long term broad band optical photometry of this 
object \cite{ba2} shows possible evidence for a variable source, with
suspicious small amplitude variability in the period 1968 
- 1986. 

The {\em HST} image shows a bright QSO with some faint nebulosity
extending approximately 2.2\arcsec\ to the north east. Three 
companion sources can be seen 4.3\arcsec\ to the south west, 12.2\arcsec\
to the south and 13.2\arcsec\ to the south east. A 
host galaxy was detected at greater than 97.5\% confidence, however
fitting a galaxy profile to the underlying light profile 
proved impossible. The measured host magnitude was also extremely large,
at $M_{I} = -27.3$. It is therefore more likely that 
we are merely detecting a profile of light from the AGN that has been scattered within
the host galaxy, rather than the host galaxy itself, hence the host is unresolved. 
The IR SED is best fit by an AGN (70\%) at 65\degr\ line of sight orientation, and a 37Myr starburst (30\%) with $\tau_{UV}=660$.  
The fit shows that the torus dominates the emission from 60$\mu$m to the UV.  The starburst dominates the 
emission longward of $60\mu$m. The orientation of the AGN model implies that the Broad Line regions should 
be visible in the optical, which is consistent with the optical morphology.

\subsubsection{LBQS1220+0939}
$z = 0.68$. This very compact QSO possesses largely symmetrical, dim,
extended nebulosity and appears to be morphologically 
undisturbed, except for some faint inhomogeneities to the south west.
There is a very dim companion source 2.1\arcsec\ to the 
south west and two brighter companions $<$ 11\arcsec\ to the north east. A
PSF + host galaxy fit is preferred to a pure PSF fit 
at $>99\%$ confidence, with an elliptical being strongly preferred to a
spiral. The host galaxy has $M_{I} = -25.0$, and a 
scalelength of $\sim$88Kpc. The average ellipticity of the source is 0.2.
There is a gradient of decreasing ellipticity 
from the central regions ($\langle\epsilon\rangle = 0.1$ at 0.25$\arcsec$)
to the outer regions ($\langle\epsilon\rangle = 0.3$ 
at 0.80$\arcsec$). There are however no large changes in ellipticity
between adjacent isophotes and there is no evidence 
for isophotal twisting, suggesing that the host is undisturbed. 
There is insufficient IR photometry to draw any conclusions about the IR 
power source.

\begin{figure}
\rotatebox{-90}{
\centering{
\scalebox{0.75}{
\includegraphics*[9,18][268,332]{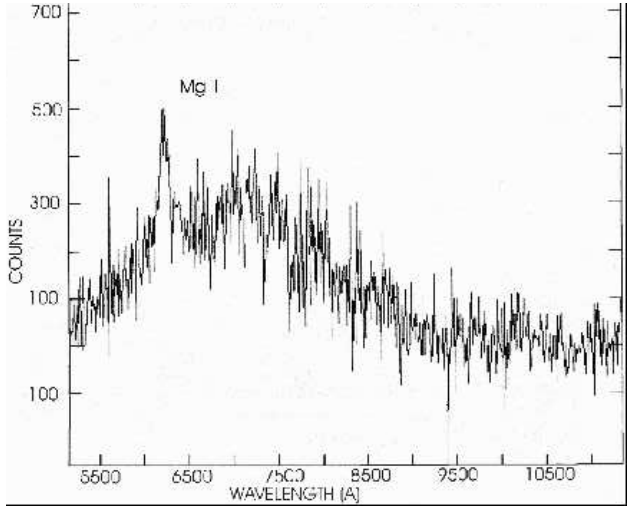}
}}}
\caption{Optical spectrum for F14218+3845, taken by Andy Lawrence with the FOS-I 
spectrograph on the Isaac Newton Telescope as part of the {\em IRAS} FSS-$z$ survey. 
\label{14218_spec}}
\end{figure}

\subsubsection{F12358+1807}
$z = 0.45$. This moderately bright QSO shows no signs of tidal
interaction, the surrounding nebulosity being extremely symmetric. 
There is however a bright spiral galaxy 8.6\arcsec\ to the north whose
magnitude ($m_{814} = 20.05$) suggests that it may be a
companion rather than a chance foreground object. The spiral appears to be
morphologically undisturbed and the spatial separation 
between it and the source is too great for direct interactions. 
The IR power source of this object cannot be constrained as there is insufficient IR 
photometry available.

The host galaxy is formally detected at $>99\%$ confidence and an
elliptical profile is strongly preferred to a spiral profile 
($\chi^{2}\sim 1.0$ vs. $\chi^{2}\sim 4.0$). Fitting elliptical isophotes
shows that the ellipticity is of the order 0.1 - 0.2. 
The host has an absolute {\em k} corrected magnitude of -24.5, and a
derived scalelength of 7.1Kpc.

\subsubsection{F12509+3122}
$z = 0.78$. This QSO is very symmetrical and shows no signs
of interaction. There is however a large, bright companion galaxy 15.5\arcsec\ 
to the south whose magnitude ($m_{I} = 17.9$) suggests physical association 
with the source rather than chance spatial coincidence. The companion appears to be
morphologically undisturbed, but has a bright nucleus. The light distribution 
of the companion is best fit by an elliptical profile, the {\em HST} image suggests 
it may be an S0 galaxy. Interestingly, this galaxy, if it is a companion, has a 
comparable magnitude ($M_{I} = -25.2$) to the QSO host galaxies, and is also similar 
to the brightest ellipticals found in moderate to rich galaxy clusters. In
addition to the bright companion there are also four other smaller, dimmer 
companion sources.

The host galaxy is formally detected with $>97.5\%$ confidence, has an absolute $k$ 
corrected magnitude of -25.2, and a scalelength of 10.3Kpc. The 60$\mu$m-25$\mu$m 
colour suggests that an AGN interpretation of the IR SED is most likely, although previous studies 
\cite{rr2} imply an unquantified, but significant starburst component.
The measured ellipticity of the host was found to be small, $\langle \epsilon \rangle = 0.15$.

\subsubsection{F14218+3845}
$z = 1.21$. Overall, the IR emission from this object is best explained as being predominantly ($74\%$) starburst 
in origin \cite{ve}. The starburst dominates only at wavelengths longward of $30\mu$m. Below $30\mu$m the AGN torus dominates. The fit 
predicts that an AGN should be responsible for most of the optical emission, which is consistent with the optical 
morphology. The optical spectrum of this source, taken as part of the {\em IRAS} FSS-$z$ 
survey \cite{ol} on the Isaac Newton Telescope using the Faint Object Spectrograph 
(FOS-I), is given in Figure \ref{14218_spec}. The single broad line is most plausibly 
interpreted as MgII. Another possibility is H$\beta$, but the absence of an H$\alpha$ line 
at longer wavelengths makes this interpretation unlikely. There are two small dim companions 
5.8\arcsec\ away to the south east and south west, and one slightly more extended companion 
13.5\arcsec\ to the south west. The source itself is very small. A host galaxy was detected 
at $>99\%$ confidence, although it proved impossible to distinguish between a de Vaucoleurs 
or spiral galaxy profile as both profiles produced an equally good fit to the data. The host 
galaxy magnitude ($M_{I} = -24.7$) is much closer in value to the total source magnitude than for 
any of the other sources. There is an extremely faint inhomogeneity about $0.5$\arcsec\ to 
the southeast in the form of a very small plume of nebulosity, this feature is however barely 
detected in the {\em HST} image ($\sim3\sigma$) and the PSF subtracted image shows no 
signs of disturbance. We thus conclude that this object is not involved in ongoing interactions. 

On {\em prima facie} grounds, the HST image of this source shows possible gravitational lensing signatures. 
Further investigation however showed no evidence for lensing. The very 
faint `ring' surrounding the QSO is at the same radius as the Airy ring
in the normalised TinyTim PSF generated for this 
object, and is therefore virtually certain to be a PSF artifact. The
likelihood of seeing a symmetrical Einstein ring with perfectly 
uniform luminosity around the entire circumference is in any case fairly
small, and would require perfect or near-perfect alignment between a 
circularly symmetric lens and an extended source. There is no tangential
stretching of the 
source, and there are no other structures or nearby companions that
suggest any plausible lensing scenario. This implies that the 
derived IR luminosity is intrinsic to the source.

\subsubsection{F15307+3252}
$z = 0.93$. This object was first identified as an HLIRG by Cutri et al
\shortcite{cu}. It was, after F10214+4724 and P09104+4109, 
the third HLIRG to be discovered, and bears many similarities to its
predecessors. Optical spectropolarimetry by Hines et al \shortcite{hi} 
shows a highly polarized continuum, and a broad emission line region that
is typical of QSO's. Nearby companions have been detected 
in {\em K} band images \cite{so1}. Liu et al \shortcite{li} discuss the
properties of this object and, using deep {\em Keck} K band 
imaging combined with $1.1\mu$m - $1.4\mu$m spectroscopy, discuss whether
this object is gravitationally lensed or a giant 
elliptical in the process of formation. Their results are inconclusive.
{\em ROSAT HRI} observations \cite{fa} show no X-ray 
emission down to a limit of $2\times10^{-4}L_{bol}$ indicating either
atypically weak X-ray emission or that little of the X-ray 
flux is scattered into our line of sight. 

CO line and rest frame 650$\mu$m observations \cite{yu} derive upper
limits to the CO molecular gas mass of 
$5\times10^{9}M_{\sun}$, and a dust mass of $0.4 -
1.5\times10^{8}M_{\sun}$. Both these values are lower than those typical
for 
the more gas rich IR galaxies. Further observations of this object show a
Seyfert 2 spectrum \cite{cu}. This was later confirmed 
by mid IR spectroscopy \cite{ev}, which also found a molecular gas mass
comparable to or less than that found in the most gas rich 
galaxies observed locally, that the relatively `warm' $60\mu$m$/100\mu$m
colours imply that most of the IR emission emanates from 
a relatively small amount of warm circumnuclear dust ($M_{dust} = 10^{5} -
10^{7} h^{-2} M_{\sun}$ which implies a molecular gas 
mass $M_{gas} = 10^{7.3} - 10^{9.3} h^{-2} M_{\sun}$), and that the narrow
emission lines and  $L_{ir}/L_{bol} = 0.90$ imply that 
the AGN is mostly obscured. The non-detection of CO poses a problem for a purely starburst
interpretation \cite{rr2}, and IR SED modelling \cite{ve} implies that an AGN is responsible 
for the majority ($68\%$) of the IR emission.

The {\em HST} image shows an interacting system with three bright
components. The most prominent source 
is large, bright, and has some small, bright arclike structures
immediately to the south west. The two smaller components are 
located 1.9\arcsec\ to the south east, with the southerly one being
noticeably brighter than the northern one. Extremely faint, 
diffuse emission is also present surrounding all three components. There
is one small dim companion and one small bright companion 
7.3\arcsec\ and 13.2\arcsec\ to the south respectively.

\begin{table}
\caption{Photometry for features within F15307+3252 \label{15307_phot}}
\begin{tabular}{@{}ccccc}
No.     & Flux (mJy)         & m$_{I}$ & M$_{I}^{1}$ & m$_{K}$$^2$ \\
{\em 1} & 5.7$\times10^{-3}$ & 18.7    & -24.8       & 17.6        \\
{\em 2} & 1.1$\times10^{-3}$ & 21.9    & -21.6       & 18.2        \\
{\em 3} & 2.6$\times10^{-4}$ & 22.2    & -21.3       & 20.1        \\
\end{tabular}

\medskip
{\em 1} refers to the large north western source, {\em 2} refers to the small south eastern source 
and {\em 3} refers to the small north eastern source. The two companion sources clearly possess different $I-K$ 
colours, ruling out a quadruple lensing scenario. $^1$No applied {\em k} correction. $^2$Taken from Liu et al 1996.
\end{table}

\begin{figure*}
\begin{minipage}{170mm}
\centering{
\epsfig{figure=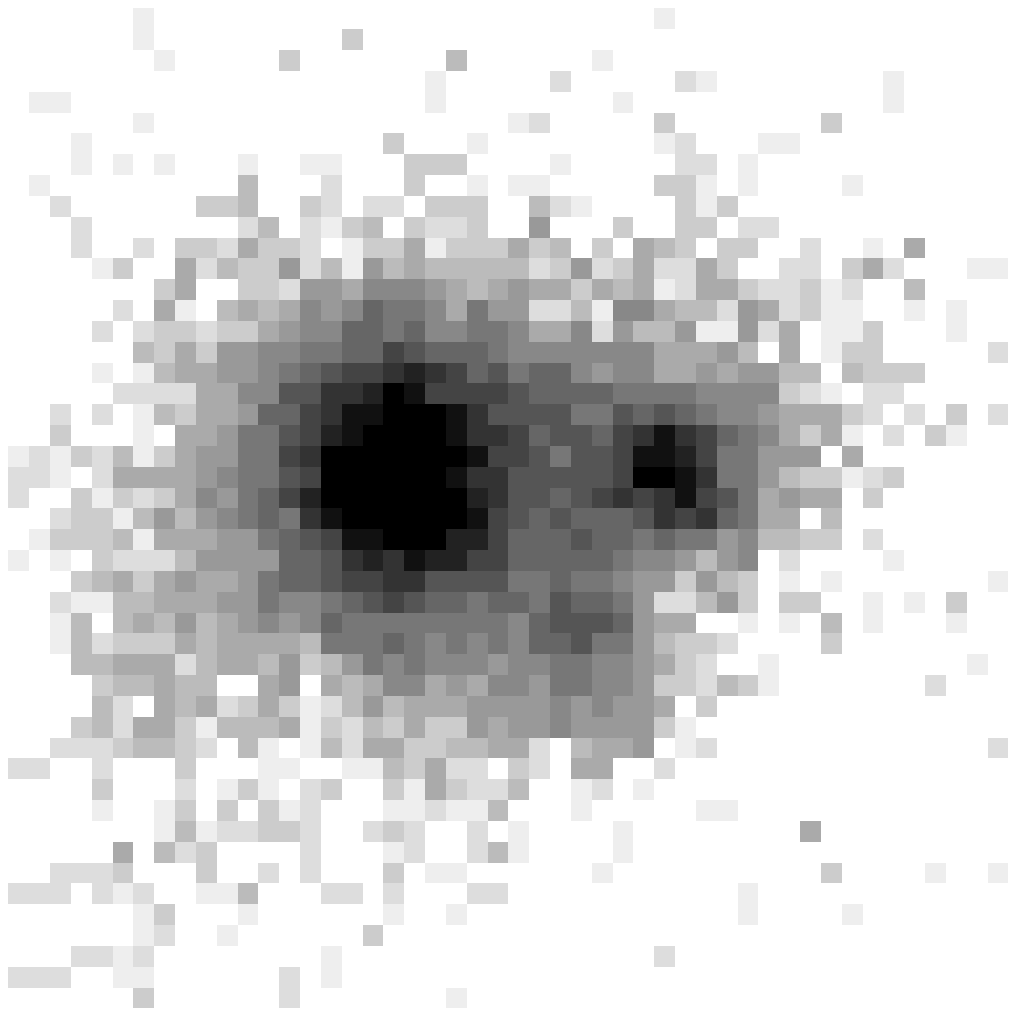,width=55mm}
\epsfig{figure=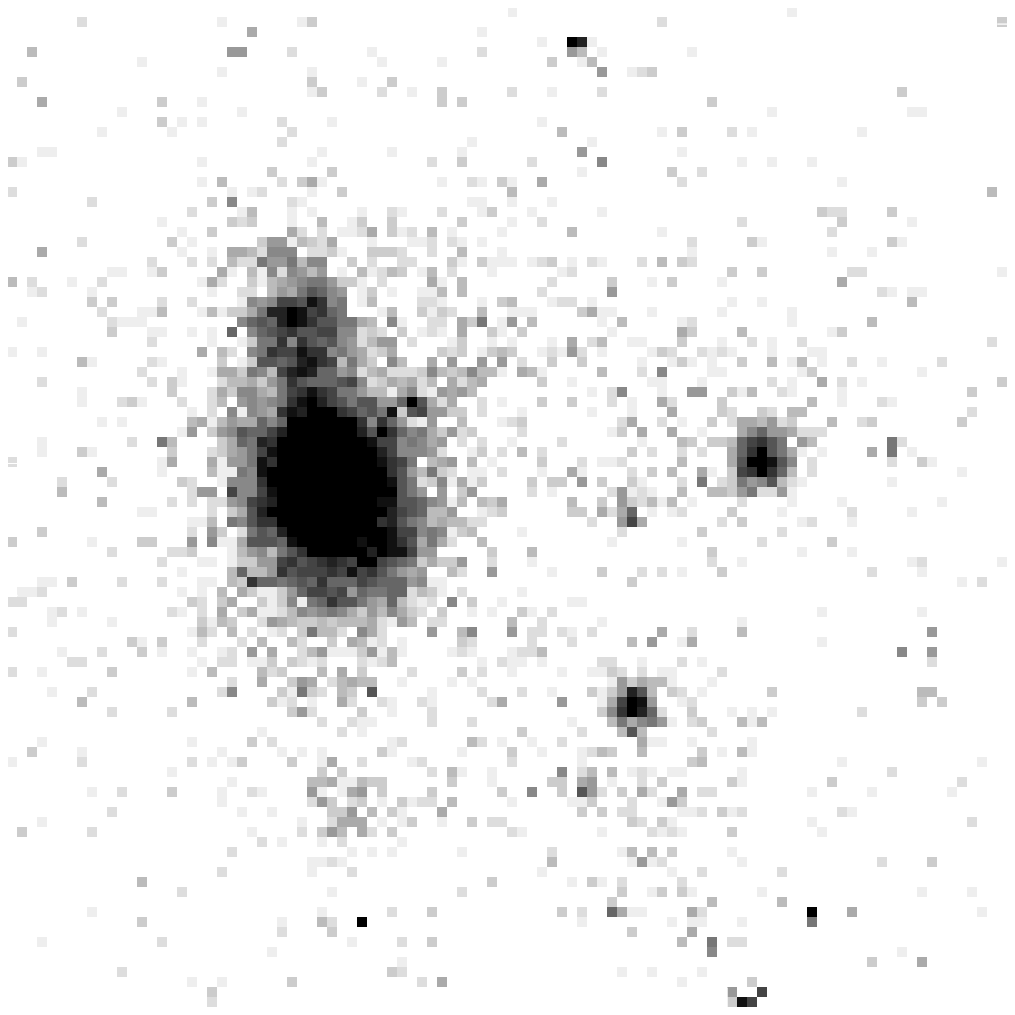,width=55mm}
}
\end{minipage}
\caption
{
{\em Keck K} band image of F15307+3252 (left, supplied by Michael Liu) and {\em HST I} band image. 
Orientation and scale are the same as in Figure \ref{hlirgs_im_src} \label{15307_keck}
}
\end{figure*}

The extremely high resolution image of this source taken by {\em HST}, in
conjunction with the {\em Keck K} band image, allows the best possible 
chance for determining whether or not this system is lensed. 
Both the {\em Keck} image (see also fig. 3 of Liu et al 1996) and the {\em HST I} 
band image are given for comparison in Figure \ref{15307_keck}. The two possible 
lensing scenarios given by Liu et al are: (1) The large, bright elliptical
component is the lensed nuclear structure of a 
$z = 0.93$ galaxy, with the brighter of the two companion sources being
the foreground lens. (2) This source is a quadruple 
image gravitational lens system. By comparing the {\em HST} image to the
{\em Keck} image (kindly provided by Michael 
Liu) it is possible to examine both these possibilities.

We first examine the possibility of a quadruple image gravitational lens
system. Firstly, there is no evidence from the {\em I} band image 
for the lens itself, although this in particular does not help to rule out
a lensing scenario.
Table \ref{15307_phot} gives the magnitudes in the {\em I} and {\em K}
bands for the three components in the system observed in both the {\em HST} and 
{\em Keck} images. The two smaller 
sources in the system possess different $I - K$ colours. Although
differential reddening through a foreground lens could account for 
colour differences between the four components the 'arclike' structures to
the south west of the large elliptical component would be 
expected to be reddened in comparison to the other three components in the
system if this were the case. Instead they are observed to 
be significantly bluer as they do not appear at all in the {\em K} band
image, but are prominent in the {\em I} band image. 
The large, bright elliptical component does not appear stretched
tangentially, but shows both radial and tangential extension. 
Explaining radial extension in a quadruple image gravitational lens is not
possible unless extremely complex lensing scenarios are 
invoked, scenarios that are not plausible from the evidence in the {\em I}
band image. Although a 
quadruple lens scenario cannot be completely ruled out, it seems
overwhelmingly unlikely. 

If the bright elliptical component of the system is assumed to be the
lensed galaxy, with the brighter of the two companion sources 
assumed to be the lens, then the lack of counter images to the south of
the lens is difficult to explain. The smaller of the two 
companion sources cannot be such a counter image as it appears on the same
side of the lens as the bright elliptical, such a scenario 
would require unrealistic amounts of cosmic shear. This rules out the
possibility that the large, bright 
elliptical component is the lensed nuclear structure of a $z = 0.93$
galaxy, with the brighter of the two companion sources 
being the foreground lens. We conclude therefore that this object is not
gravitationally lensed and that the IR emission is intrinsic to the 
source.

Fitting elliptical isophotes to the largest component in this source shows
that the ellipticity varies $0.25 < \epsilon < 0.4$ as 
the radius $0.2 < r < 0.32$ in arcseconds. The best fit profile of the
large bright component is found to be an elliptical, with 
an associated scalelength of 12.4Kpc and an absolute {\em k} corrected
magnitude of $M_{I} = -26.4$.

\subsubsection{PG1634+706}
$z = 1.33$. This radio quiet object is the most IR luminous optically
selected QSO in the {\em IRAS} catalogues. The 
extraordinarily high disk luminosity ($\sim10^{40} W s^{-1}$) requires a
central black hole of $10^{10}M_{\sun}$, 
even at the Eddington accretion limit \cite{ra}. An AGN dust torus model
is consistent 
with both the {\em ISO} \cite{haa} and {\em IRAS} data, with a limit set
on any starburst contribution, a 
limit which is confirmed by the non-detection of CO \cite{ev}. The near IR
spectrum \cite{ev} 
is not consistent with a Seyfert 2 galaxy, and the derived molecular gas
mass is comparable to or less than that 
found in the most gas rich galaxies observed locally. They also find that
the relatively `warm' $60\mu$m/$100\mu$m 
colours imply that most of the IR emission emanates from a relatively
small amount of warm circumnuclear dust 
($M_{dust} = 10^{5} - 10^{7} h^{-2} M_{\sun}$, implying a molecular gas
mass $M_{gas} = 10^{7.3} - 10^{9.3} 
h^{-2} M_{\sun}$), and that the broad emission lines and $L_{ir}/L_{bol} =
0.35$ suggests that most of the AGN is 
exposed. ASCA spectroscopy of this object \cite{na} agrees with these
conclusions to some extent. Both 
the FeK$\alpha$ line and `hard tail' are not present in the spectrum,
features which are almost universal in Seyfert 2 
galaxies. The accretion disk is thus concluded to be either highly ionised
or optically thin, both of which would 
indicate that accretion is occurring close to or at the Eddington
accretion limit. An optically thick molecular 
torus is concluded to be either absent or subtending a small angle to the
X-ray source. ROSAT PSPC observations 
\cite{ra} show this source to have a steep soft X-ray slope, a natural
explanation of which is that the 
soft X-ray emission is of unsaturated comptonisation origin, covered by a
flat nonthermal X-ray component 
above $1.4 \pm 1.1$ KeV. Barvainis, Lonsdale \& Antonucci \shortcite{ba3}
confirm previous radio observations of this source which show a high 
frequency excess, attributed to a compact synchrotron source most commonly
found in radio loud objects. 

The {\em HST} image shows a large, bright, symmetrical source with no
surrounding nebulosity. There are no companions within the 36\arcsec\ field of view. 
A host galaxy was detected for this object, and a spiral disk profile is 
preferred over an elliptical profile. The host was however only detected at 
$\sim80$\% confidence over a pure PSF fit, and the derived host galaxy 
magnitude, at $M_{I}=-29.4$, is also unphysical. It is therefore more likely, 
as in the case of PG1148+549 that we are merely detecting a profile of scattered 
light from the AGN and that the host galaxy is unresolved. 
The IR SED is dominated by the AGN torus component (83\%), with line of sight
inclination of 90\degr. The starburst
model of age 6.6Myr and $\tau_{UV}=220$ is required only to
explain the emission longward of $60\mu$m.

\section{Discussion}
\subsection{The Sources}
\subsubsection{Morphology}
On an {\em ad hoc} basis the sample can be divided into two groups on the basis of their morphology, namely an 
`interacting' group and a `quasar' group. The `interacting' group comprises three objects; F00235+1024 and 
F15307+3252 are both strongly interacting, F10026+4949 shows slightly weaker signs of interaction but has 
numerous nearby ($<$5\arcsec) companions. The `quasar' group comprises the other six sources, five of which are optically bright 
QSOs and one of which, F14218+3845, is a QSO where the point source is relatively underluminous.

There is a strong morphological resemblance between the `interacting' group of HLIRGs and previous {\em HST} imaging of 
ULIRGs \cite{su,bor}. The strong resemblance between three sources in this sample with previously observed interacting 
ULIRGs, and that the rest of the sources are QSOs, raises the question as to whether any of the QSOs in the sample have 
arisen from interactions or mergers. It has been suggested \cite{sa2} that ULIRGs represent a transition phase between 
galaxy mergers and optically luminous quasars. The formation of quasar nuclei in ULIRGs has been investigated by Taniguchi, 
Ikeuchi \& Shioya \shortcite{ta}. They conclude that a supermassive black hole (SMBH) of mass $\geq10^{8}M_{\sun}$ can form 
if either one of the progenitor galaxies contains a `seed' SMBH of mass $\geq10^{7}M_{\sun}$ undergoing efficient Bondi type 
gas accretion during the merger, or by collapse of some star cluster containing compact remnants (black holes or neutron stars) 
on a time scale $\sim 10^{9}$ years. A more recent study of ULIRGs \cite{far} using HST, proposes that ULIRGs as a class are not simply 
a transition stage between galaxy mergers and QSOs, but instead are a much more diverse population whose evolution is driven 
solely by the morphologies of the merger progenitors and the local environment.

\begin{figure}
\rotatebox{90}{
\centering{
\scalebox{0.36}{
\includegraphics*[54,54][558,718]{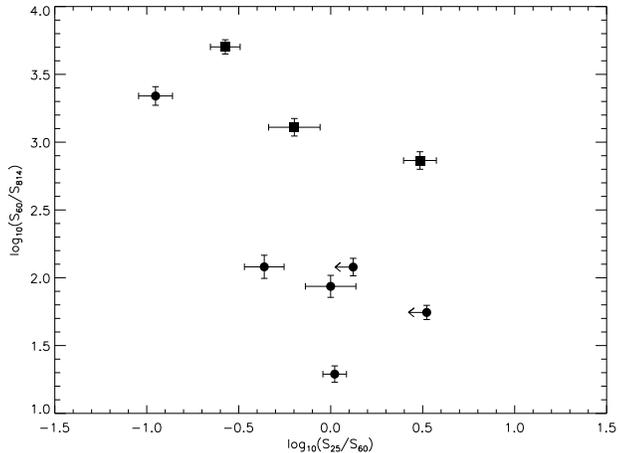}
}}}
\caption{Colour - colour plot for the 9 sources in the sample. K corrections have been applied to the {\em I} 
band and IR fluxes so that the plot approximates rest-frame colours. The circles are those sources classified as 
QSOs on the basis of their optical spectra, and the squares are the interacting sources.
\label{colcolplot}}
\end{figure}

Signs of interaction in the immediate vicinity of the QSOs in the sample would be hard to detect due to the bright central 
source, and the relatively high redshifts of the objects. A significant fraction of ULIRGs in which QSO activity is apparent 
have shown clear signs of interactions \cite{le,cl}, and observations of QSOs from the PG quasar survey \cite{cl2} showed 
that QSOs with disturbed hosts were significantly more likely to have a higher FIR luminosity than those QSOs with undisturbed 
hosts. These studies have however involved objects at redshifts $\leq 
0.2$. Of the six QSOs in this sample, three show faint signs of morphological disturbance, two are apparently undisturbed and one 
(PG1634+706) is too distant to tell. Given that all the objects that do not have an optical QSO show clear signs of interaction, 
and that the other six are QSOs it seems plausible that the QSOs in the sample could represent the last stage in a merger event. 
HLIRGs would then be plausibly interpreted as a simple extrapolation of the ULIRG population to extreme luminosities. 
From the {\em HST} and IR data alone it is however not possible to rule out that some or all of the QSOs are not a simple 
extrapolation of the ULIRG population to extreme luminosities and that the IR emission is not triggered by interactions.

\subsubsection{An Optical/IR Comparison}
The IR power sources for those objects with previously modelled $1-1000\mu$m IR SEDs can be found in Table \ref{hstobs}. 
For the five sources with sufficiently comprehensive IR photometry to allow detailed modelling of the IR SED, two are starburst 
dominated and three are AGN dominated, although all five require both a starburst and an AGN to explain the IR emission. Two 
sources with less comprehensive IR photometry which cannot be modelled using detailed SED models show evidence for an AGN dominating the 
IR emission, with a smaller, but unquantified contribution 
from a starburst. The remaining two sources have an unknown IR power source as they have only one IR data point. 

Table \ref{hstobs} gives a direct comparison between the {\em HST} morphology and the best fit model to the IR SED. Of the two 
sources that are starburst dominated, one (F00235+1024) is interacting and the other (F14218+3845) is a QSO. For the five systems 
where an AGN is the dominant IR power source, two (F10026+4949 \& F15307+3252) are strongly interacting and the rest are QSOs.
It can be argued that all the starburst systems are triggered by interactions, one is in ongoing interactions, whilst the 
second (F14218+3845) shows no clear signs of interaction but does possess close companions. For the five 
AGN dominated systems however, two are in ongoing interactions, but three are optically luminous QSOs with no signs of
interaction or very close companions. There is therefore no clear correlation between the IR power source and the optical morphology 
in this sample of HLIRGs. The number of sources considered here is however too small to draw truly robust statistical
constraints on the HLIRG population. This means that these conclusions are by nature quite speculative, and should be regarded 
with some caution.

\begin{figure*}
\begin{minipage}{150mm}
\epsfig{figure=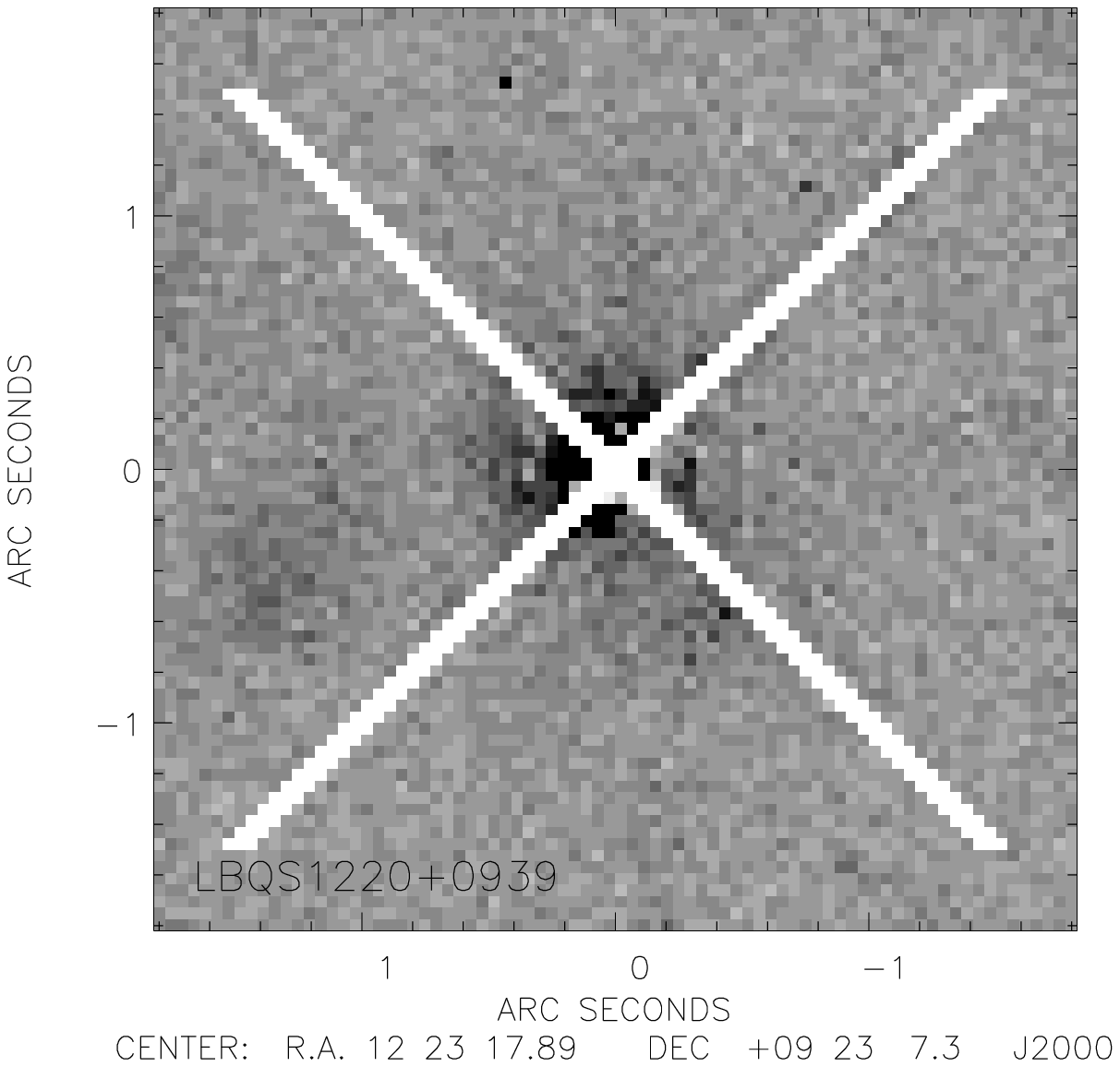,width=70mm}
\epsfig{figure=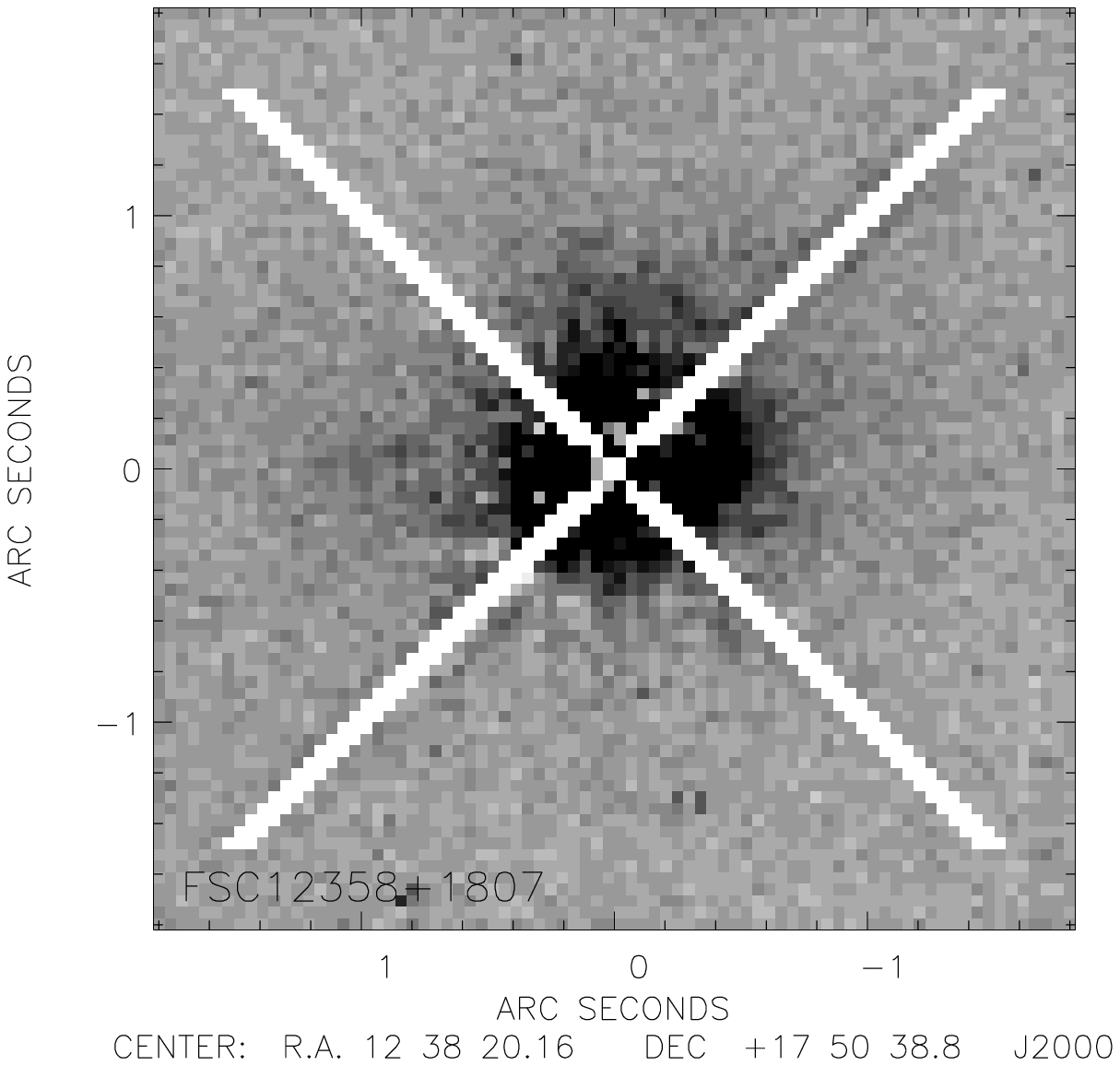,width=70mm}
\end{minipage}
\begin{minipage}{150mm}
\epsfig{figure=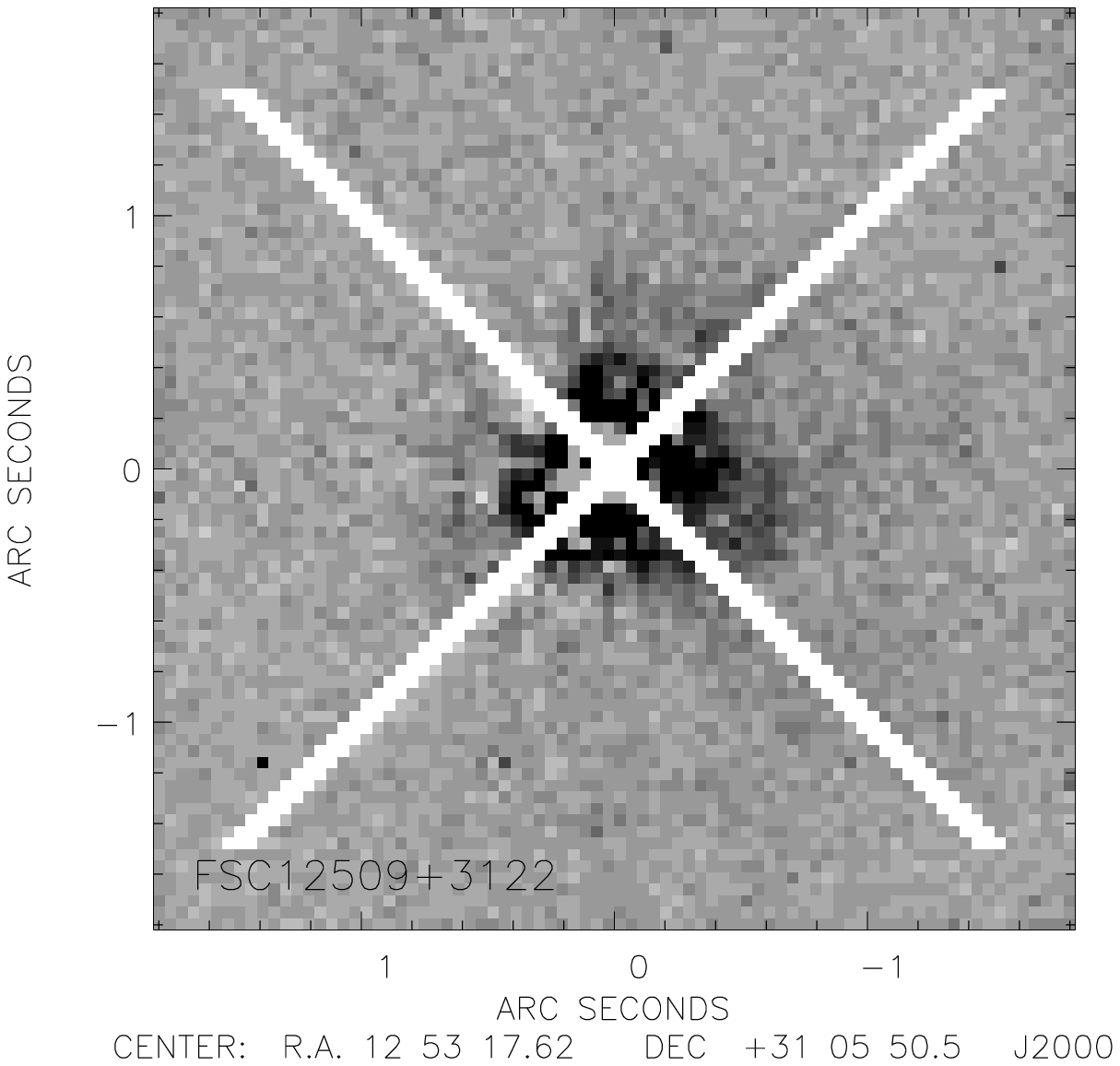,width=70mm}
\epsfig{figure=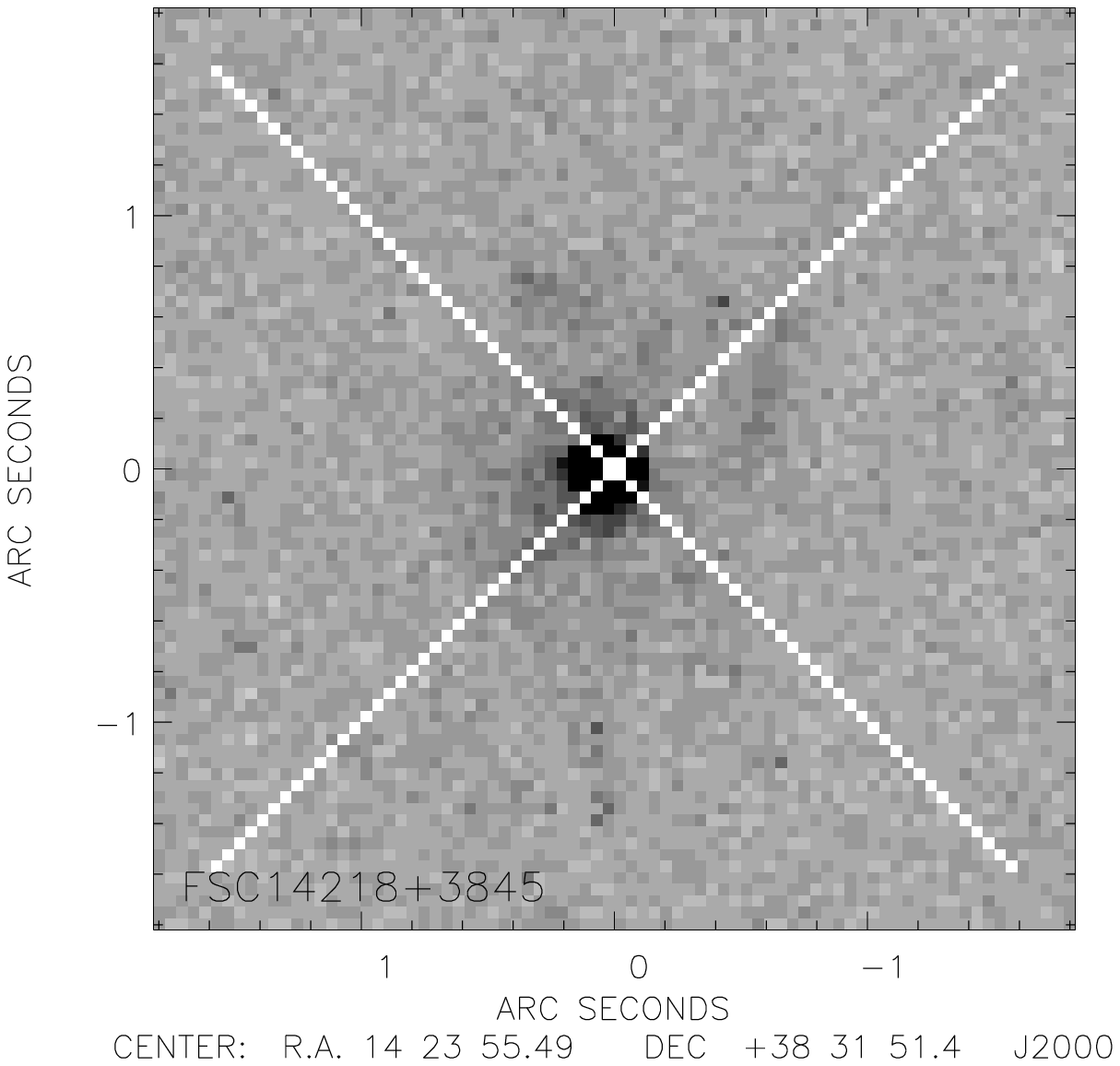,width=70mm}
\end{minipage}
\caption{{\em HST} F814W images of the four resolved HLIRG QSO host galaxies. Residual PSF features have been masked 
out. \label{hlirgs_im_host}}
\end{figure*}

For the five sources with well sampled SEDs, the optical morphology correlates well with that predicted by the dominant contributor 
to the SED in the optical. For PG1148+549, F14218+3845 \cite{ve} and PG1634+706 the SED fit predicts that the optical emission is 
dominated by an AGN, which is in excellent agreement with the optical images. For F00235+1024 and F15307+3252 \cite{ve} the SED fit 
predicts that the AGN does not contribute significantly to the optical emission. Although the unquantified contribution from old 
stellar populations to the {\em I} band flux means that the optical data 
cannot be used to constrain the properties of the starburst, the absence of AGN both in the optical images and in the optical region 
of the SED are consistent with each other. 

A colour-colour plot ($log (S_{25}/S_{60})$ vs. $log (S_{60}/S_{814})$) for the nine sources in the sample can be found in Figure 
\ref{colcolplot}. All of the optically luminous QSOs are localised in the bottom right hand corner of the colour-colour plane. 
The three interacting sources are also localised in the $log (S_{60}/S_{814})$ axis, but are more spread out along the $log (S_{25}/S_{60})$ axis
due to their varied AGN fractions. F14218+3845 is both substantially less luminous in the optical than the other QSOs in the sample (probably due
to higher levels of obscuration in the nuclear regions), and contains a starburst component contributing dominantly to the IR emission 
longward of $60\mu$m \cite{ve}, hence its position in the top left of the colour-colour plane close to the interacting sources.

\subsubsection{Gravitational Lensing}
The extraordinarily high bolometric IR luminosities observed in these objects suggests that 
magnification via gravitational lensing may play a significant role. Gravitational lensing has been 
observed previously in four IR selected hyperluminous sources, these include the `Cloverleaf' quasar 
and F10214+4724. Signs of lensing were therefore a primary search objective for this sample. One of
the sources was known to be a strong candidate for lensing, as morphological information for 
F15307+3252 from ground based imaging suggested that it may be a lensed system of some 
kind \cite{li}.

It came as quite a surprise therefore, when none of the sources in the sample showed any evidence 
for gravitational lensing. The previous observations of 
HLIRGs that have detected lensing using {\em HST} indicate that this phenomenon is at least present 
to some esxtent in the HLIRG population. Including our sample suggests that only a small minority (around 15\% - 
20\%) of HLIRGs have been mistakenly classified as such due to the effects of lensing. This 
percentage should be treated with a degree of caution; the sample presented here is at a lower 
median redshift than that of previously confirmed lensed sources, the sample of {\em HST} imaged HLIRGs 
is also as a whole not homogenous. There is also the possibility that these objects may be lensed by a 
foreground cluster, such a lensing scenario might be expected to produce weaker signs of lensing than 
would be observed if any of these objects were being lensed by a single source. This possibility can 
only be fully examined by a quantitative study of the environments of these objects, although the 
complete lack of lensing signatures in any of the objects makes a cluster lens scenario unlikely.

\subsection{The Host Galaxies}
Previous surveys of QSOs using {\em HST} have generally been extremely successful in detecting host galaxies. 
Urry et al \shortcite{ur}, give results for a sample of BL Lacertae objects, detecting host galaxies out to a redshift of 
$z = 0.7$. They find that the hosts are all extremely luminous ellipticals with $\langle M_{I} \rangle = -24.6$. The derived 
ellipticities are generally low and morphologies are smooth. Imaging of a sample of 14 quasars (6 radio loud, 5 
radio quiet, 3 radio quiet ultraluminous infrared quasars) by Boyce et al \shortcite{bo} showed that all the radio loud 
quasars had elliptical hosts, the radio quiet quasars possessed both elliptical and spiral hosts, and that the 
radio quiet ultraluminous infrared quasars lay in violently interacting systems. Host galaxies were on average 
0.8 mag brighter than $L^{*}$. {\em K} band imaging of of a sample of 14 luminous radio quiet QSOs \cite{per} found that 
two of the host galaxies were violently interacting, and the rest were undisturbed ellipticals.
Results from imaging of two luminous radio quiet quasars \cite{ba1} show that both quasars 
reside in apparently normal host galaxies, one being an elliptical and one a spiral. Disney et al \shortcite{di} present 
{\em HST} imaging for a sample of four QSOs, finding that all four have elliptical hosts. Although they find that the 
morphologies of the hosts are all featureless they argue that the presence of multiple very close companions is evidence 
for interactions as the trigger for the quasar activity.  

We resolved host galaxies in seven of the nine objects, with the host morphology resolved in five of those cases.
Host properties were reliably determined up to a redshift of 0.78, except in the case of F14218+3845 where the 
relatively underluminous point source allowed determination of the host galaxy magnitude, but not the profile. The 
host galaxy was not resolved for the two PG quasars, probably because the extremely bright central source in these 
two objects creates a profile of scattered light which completely masks the host galaxy. The morphology of F00235+1024 was too 
disturbed to fit a galaxy profile to. 

The five reliably resolved profiles were all best fitted by an elliptical profile with at least 
97.5$\%$ confidence. The average QSO host galaxy magnitude was found to be $\langle M_{I} \rangle = -24.9$, more than 1.7 magnitudes 
brighter than $L^{*}$. For those sources that do not display a point source the galaxies were brighter, with $\langle M_{I} \rangle = -25.5 
\pm0.9$. The derived scalelengths were also large, all but one falling in the range $6.7 < r_{e} < 12.5$. 
The scalelength for  LBQS1220+0939, at $\sim 88$Kpc is extremely large, and is comparable to the largest observed 
ellipticals. The possibility exists that this value is erroneous, possibly due to AGN light scattered within the host galaxy 
or (less likely) a poorly subtracted PSF, however the faint, extended emission surrounding this source is supportive of a large 
scalelength.

\begin{figure}
\rotatebox{0}{
\centering{
\scalebox{0.58}{
\includegraphics*[10,10][403,360]{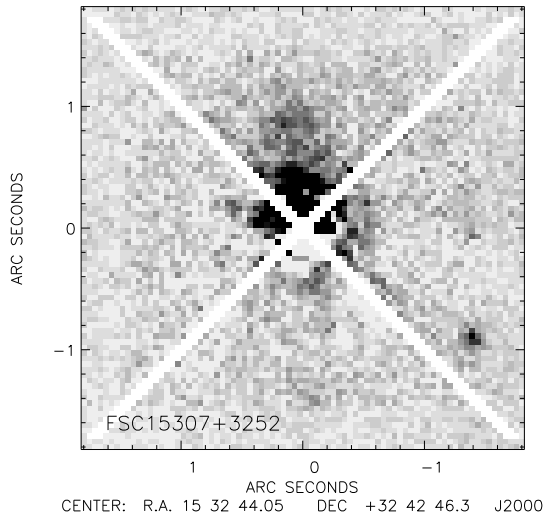}
}}}
\caption{A simulation in which an observed, saturated Planetary Camera {\em I} band PSF is added to the large elliptical component of 
F15307+3252 shown in Figure \ref{hlirgs_im_src}. A theoretical TinyTim PSF has then been subtracted off, following the procedures 
described in \S 3.2. Residual PSF features have been masked out. The weak signs of morphological disturbance seen in the original 
image are still clearly visible. The less luminous of the two companion sources is visible on the lower right of the image. \label{demo_psfsub}}
\end{figure}

We performed simulations to quantify the efficiency of host galaxy detection, and to determine whether 
signs of morphological disturbance would be resolved after PSF subtraction. In order to make the simulations realistic 
we took the large bright elliptical component of 
F15307+3252, which shows weak signs of morphological disturbance, and added on an observed saturated PSF taken with the 
F814W filter from the {\em HST} PSF archive. TinyTim PSFs were then generated 
for the position of this observed PSF. Subtraction and fitting procedures were carried out as for the QSOs in the sample. 
A host galaxy was detected at $>97.5$\% confidence, and an elliptical profile was clearly preferred to a spiral profile.
The measured host galaxy magnitude was found to be $M_{I} = -24.9 \pm 0.2$, with an associated scalelength of 
$r_{e} = 12.8\pm 0.8$ Kpc. The $1\sigma$ errors of both these values lie within the original derived values for this 
object. Figure \ref{demo_psfsub} shows the galaxy after PSF subtraction. The signs of disturbance in the original image 
are still clearly visible. We therefore conclude that, had any of the detected host galaxies of the QSOs shown signs of 
interaction of a level comparable to that seen in the large elliptical component of F15307+3252, we would have 
detected them.

The derived host galaxy properties are similar to the properties of the hosts of BL Lac 
objects observed by Urry et al. They are also similar to the host galaxy properties derived for radio quiet QSOs 
by Boyce et al and Bahcall et al, although we do not find any spiral hosts. Observations of the brightest ellipticals in the Virgo 
and Coma clusters \cite{ca,jo} show that they have very similar properties to our host galaxies. Interestingly, both 
{\em IRAS} quasars imaged by Boyce et al had host galaxies classified as interacting. If all the QSOs in our sample 
arose due to interactions and mergers it would not be unreasonable to expect some of the hosts in our sample to be 
interacting systems, although the combination of a luminous point source and relatively high redshift may well serve to mask any such 
features. The low measured ellipticities for the QSO hosts suggests that the hosts are either featureless or in the very last stages 
of a merger. As ULIRGs are thought to be mergers between two or more gas rich spiral galaxies, and as a merger between two 
such galaxies is thought to form an elliptical galaxy \cite{bar}, the properties of our host galaxies are consistent with 
being ULIRGs in the last stages of a merger.

\subsection{Companion Sources}
It has long been suspected that IR emission in ULIRGs is triggered by interactions and mergers \cite{sa1}, one possibility  
\cite{bor} is that the IR emission may arise due to multiple mergers and interactions in Hickson Compact 
Groups. Results from the analysis of the QSO host galaxies suggest that HLIRGs may reside in rich clusters, given that the 
hosts are very similar to the brightest cluster ellipticals observed locally.
All but one of the sample show at least one companion source within 36\arcsec. Two sources have single, large, nearby bright 
companions, one of which is a very bright elliptical, the other a spiral. The other six all show at least two 
smaller, dimmer companions within a 16\arcsec\ radius. The one exception, PG1634+706, apparently has no companions within 0.5Mpc, 
suggesting that this source does not reside in a cluster. 

These results, even when taken in conjunction with the results for the host galaxies, do not provide conclusive evidence 
for or against a cluster environment for all HLIRGs, but do suggest that some HLIRGs may reside in a rich environment. 
The environments of the HLIRGs in this sample will be presented in more detail in a future paper.

Studies of the environments of ULIRGs \cite{bor0,sa0} have shown that these objects do not appear to reside in regions of 
higher than average density. Most ULIRGs appear either isolated or in systems similar to the local group. Results from the 
{\em IRAS} surveys have shown that most ULIRGs are strong interactions between gas rich spirals with a relative mean 
velocity of $\leq 200$km s$^{-1}$. These are not typical conditions of rich cluster environments, where relative mean 
velocities are much higher and many galaxies are either gas poor spirals, or ellipticals \cite{sa1}. A study of the cluster 
environments of radio loud and radio quiet AGN has been performed by Mclure \& Dunlop \shortcite{mcl}. They showed that both 
classes of object resided 
in rich environments of approximately Abell class 0, although there was a large scatter. They also found no evidence 
for an epoch dependent change in the environments, and that AGN cluster environments as a whole were consistent with 
being drawn at random from the general cluster population.

The tentative hypothesis presented here that some fraction of HLIRGs may reside in a rich cluster 
environment, if confirmed, would be an important result and would raise problems for interpreting 
HLIRGs as a simple extrapolation of the {\em IRAS} galaxy population to extreme luminosities. 
Density evolution with redshift cannot explain sparse environments at $z \sim 0.2$ and rich 
environments at higher redshift. In any case strong luminosity evolution, as opposed to density 
evolution, is most appropriate for explaining faint radio source counts \cite{rr1}. HLIRGs may be 
the cluster analogues of ULIRGs; the higher relative mean velocities found in rich clusters would 
make the merger driven IR luminous phase shorter, this would also provide a formation mechanism for 
cluster elliptical galaxies. This scenario would imply some correlation between luminosity and 
envioronment in the more luminous of the {\em IRAS} galaxies. Alternatively HLIRGs may be an entirely 
separate class of object.

\section{Conclusions}
We have performed {\em HST} {\em I} band imaging for a sample of nine Hyperluminous Infrared Galaxies. 
The results are:

(a) The morphologies in the sample include both strongly interacting systems, and QSOs with no clear signs of 
ongoing interaction.

(b) No source in the sample shows any evidence for gravitational lensing. This includes F15307+3252, for which lensing had 
been previously suspected. Lensing 
is thus concluded to be entirely absent from the sample or at such a low level that it has a negligible impact on the observed 
luminosities.

(c) The QSO host galaxies are all extremely luminous ellipticals with properties comparable to the brightest 
cluster elliptical galaxies observed locally.

(d) There is no clear correlation between the IR power source and the optical morphologies. Of the seven sources with enough IR 
photometry to investigate the IR power source, two are starburst dominated and five are AGN dominated, although all seven show 
evidence for some level of AGN emission. One of the starburst dominated systems is in ongoing interactions, the other is a QSO 
with close companions. Of the five AGN dominated systems, two are interacting and three are QSOs with no signs of interaction. 
There is however good agreement between the optical morphologies and the dominant contributor to the optical emission predicted 
by the SED. 

(d) Two of the sources in the sample show a single bright companion, six show at least two small dim companions. One source is 
very probably isolated

Previous observations of HLIRGs have seen evidence for amplification of intrinsic luminosities via gravitational lensing. None 
of our sources however showed any evidence for this phenomenon, including one for which lensing had been previously strongly suspected. 
We conclude that only a small minority (perhaps 15\% - 20\%) of HLIRGs have been misclassified due to gravitational lensing. 

These results are consistent with the majority of the HLIRG population constituting the extremely 
bright end of the ULIRG population, i.e. where interaction induced star formation and quasar activity lead to the 
very high IR luminosities. The evidence for this can be summarised as; (1) Three of the sources lie in interacting 
systems, which bear a strong morphological resemblance to ULIRGs (2) For the five sources with sufficient data to 
accurately model the IR SED, the IR emission is best explained as arising from a mixture of starburst and AGN activity, 
which is similar to that found amongst ULIRGs. (3) The QSO host galaxy properties correspond closely to locally observed bright 
cluster ellipticals, a cluster environment would provide a convenient source for interactions and mergers. (4) Most of 
the sources show at least two nearby companions. (5) One of the bright companion sources is a galaxy with similar 
properties to bright cluster ellipticals. 

There are however some important points to note about these conclusions. None of the QSOs or QSO hosts show any clear signs of 
interaction, signs which have been observed in ULIRGs in which QSO activity is apparent. Such signs may however be hidden due to 
high redshift, luminous central source, or because hyperluminous activity takes place in mid to late stage mergers where the signs 
of merging are diminished. The evidence that the QSOs reside in luminous elliptical hosts is only suggestive that some may also 
reside in rich cluster environments, and is not conclusive. For example one source (PG1634+706) is apparently isolated. Further 
analysis is required to make more concrete statements about the local environments of HLIRGs. If the tentative suggestion 
put forward here that some HLIRGs may reside in a rich cluster environment is found to be correct, then this would pose problems 
for HLIRGs being a simple extrapolation of ULIRGs to extreme luminosities as ULIRGs are known not to reside in rich clusters. It 
is thus possible that some HLIRGs are the cluster analogues of ULIRGs, or that the IR emission in some HLIRGs may not be triggered 
by interactions.

It must be emphasized that these conclusions are based on a relatively small sample of objects, essentially 
those HLIRGs that have been observed at sufficient resolution to reliably determine morphologies, host galaxy 
properties and lensing characteristics. It is not possible to draw statistically significant conclusions from a sample of this size. 
Observations of a larger number of HLIRGs to the highest possible depth and resolution are the only way to confirm or refute the 
conclusions drawn above.

\section{Acknowledgments}
We would like to thank Stephen Serjeant and John Krist for invaluable advice on PSF 
subtraction, Stephen Warren for advice both on PSF subtraction and gravitational lensing, John Biretta 
and Stefano Casertano for assistance with data reduction and analysis, Andreas Efstathiou 
for helpful discussion and for providing the IR SED models, and Michael Liu for providing the {\em Keck} image of 
F15307+3252. 

The data presented here were obtained using the NASA/ESA {\em Hubble Space Telescope}, obtained at the 
Space Telescope Science Institute, which is operated by the Association of Universities for Research in 
Astronomy, Inc., under NASA contract NAS 5-2655. The work presented has made use of the NASA/IPAC 
Extragalactic Database (NED), which is operated by the Jet Propulsion Laboratory under contract with 
NASA. The Isaac Newton Telescope is operated on the island of La Palma by the Isaac Newton Group in the 
Spanish Observatorio del Roque de los Muchachos of the Instituto de Astrofisica de Canarias. D.G.F would 
like to acknowledge the award of tuition fees and maintenance grant provided by the Particle 
Physics and Astronomy Research Council. This work was in part supported by PPARC grant no. GR /K98728.

\bsp 

\label{lastpage}

\end{document}